\documentclass[12pt]{article}
\usepackage{eqsection,amsfonts}
\usepackage{latexsym,epsf,cite}

\def\thefootnote{\fnsymbol{footnote}}

\newcommand{\eq}{\begin{equation}}
\newcommand{\en}{\end{equation}}
\newcommand{\eqa}{\begin{eqnarray}}
\newcommand{\ena}{\end{eqnarray}}

\newcommand{\br}{\langle}
\newcommand{\kt}{\rangle}

\newcommand{\be}{\begin{equation}}
\newcommand{\ee}{\end{equation}}

\newcommand{\reff}[1]{(\ref{#1})}
\def\sZ{\mathbb{Z}}
\def\sR{\hbox{{\rm I}\kern-.2em\hbox{\rm R}}} 
\def\sN{\hbox{{\rm I}\kern-.2em\hbox{\rm N}}} 
\def\smfrac#1#2{{\textstyle\frac{#1}{#2}}}

\def\spose#1{\hbox to 0pt{#1\hss}}
\def\ltapprox{\mathrel{\spose{\lower 3pt\hbox{$\mathchar"218$}}
 \raise 2.0pt\hbox{$\mathchar"13C$}}}
\def\gtapprox{\mathrel{\spose{\lower 3pt\hbox{$\mathchar"218$}}
 \raise 2.0pt\hbox{$\mathchar"13E$}}}

\begin{document}
\begin{titlepage}
\vskip0.5cm
\begin{flushright}
DFTT 17/2001\\
DESY 01-074\\
IFUP-TH 99/2001\\
Roma1-1963/01\\
\end{flushright}
\vskip0.5cm
\begin{center}
{\Large\bf Irrelevant operators in the}
\vskip 0.3cm
{\Large\bf two-dimensional Ising model}
\end{center}
\vskip 1.3cm
\centerline{
Michele Caselle$^a$, Martin Hasenbusch$^b$, Andrea Pelissetto$^c$ and 
Ettore Vicari$^d$}

 \vskip 0.4cm
 \centerline{\sl  $^a$ Dipartimento di Fisica Teorica dell'Universit\`a di
 Torino and I.N.F.N., I-10125 Torino, Italy}
 \centerline{\sl $^b$ NIC/DESY Zeuthen, Platanenallee 6, 
 D-15738 Zeuthen, Germany}
 \centerline{\sl  $^c$ Dipartimento di Fisica dell'Universit\`a di Roma I
 and I.N.F.N., I-00185 Roma, Italy}
 \centerline{\sl  $^d$ Dipartimento di Fisica dell'Universit\`a di Pisa 
 and I.N.F.N., I-56127 Pisa, Italy}
 \vskip 0.2truecm
 \centerline{E-mail: Caselle@to.infn.it, Martin.Hasenbusch@desy.de,}
 \centerline{Andrea.Pelissetto@roma1.infn.it, Vicari@df.unipi.it}
 \vskip 1.cm

\begin{abstract}
By using conformal-field theory, we classify the possible irrelevant
operators for the Ising model with nearest-neighbor interactions 
on the square and triangular lattices. 
We analyze the existing results for the free energy and its derivatives 
and for the correlation length, showing that they are in agreement 
with the conformal-field theory predictions. Moreover, these results 
imply that the nonlinear scaling field of the energy-momentum tensor
vanishes at the critical point. Several other peculiar cancellations are 
explained in terms of a number of general conjectures. We show that all
existing results on the square and triangular lattice are consistent
with the assumption that only nonzero spin operators are present.
\end{abstract}
\end{titlepage}

\setcounter{footnote}{0}
\def\thefootnote{\arabic{footnote}}

\section{Introduction} \label{sec1}

The role of the irrelevant operators in the two-dimensional Ising model
with nearest-neighbor interactions
has been extensively discussed in the literature. The first important result
is due to Aharony and Fisher~\cite{af}, who showed, by using the exact
results for the free energy and the magnetization in infinite volume,
that the first correction to the susceptibility
could be explained in terms of purely analytic corrections, i.e.
without introducing any contribution due to irrelevant operators. 
The conclusions of Aharony and Fisher were strengthened by the 
analysis of \cite{gc}, that showed that the behavior 
of $\chi$ up to $O(t^4)$ was fully compatible with the 
absence of irrelevant operators.\footnote{We should also mention that recently 
a similarly unexpected cancellation was found in the 
free energy on the critical isotherm $T = T_c$ \cite{CaHa99}. }
These results gave rise to the idea (which has never
received the status of an explicit conjecture as far as we know, 
but which has been
commonly accepted in the statistical-mechanics community) that no contribution 
from irrelevant operators is present in the free energy of the 
two-dimensional Ising model with nearest-neighbor interactions. 
Of course, such a statement cannot 
be generically correct, since the lattice Ising model 
shows explicit violations of rotational invariance that 
{\em must} be due to nonrotationally invariant irrelevant 
operators. In particular, in \cite{CPRV-98}, from the analysis of the 
mass gap, irrelevant corrections with renormalization-group 
(RG) dimension $y=-2$ (respectively $y=-4$) were clearly 
identified on the square (resp. triangular) lattice. Of course, 
the question remained if these operators did contribute to the free 
energy.  

The analysis of the susceptibility of \cite{gc} has been recently 
extended in \cite{n99,g2000}.
In~\cite{g2000}, thanks to an
impressive progress in the construction and analysis of the series expansions
for the susceptibility, it was clearly shown that at least two irrelevant 
operators contribute to the expansion of the susceptibility for $h=0$
near the critical point.
However, while these results show without doubts 
the presence of irrelevant operators, they do not characterize them.
In particular, the identification of these irrelevant
operators with the corresponding
quasiprimary fields of the Ising Conformal Field Theory (CFT)
is still an open problem.  In this paper we try
to make some progress in this direction.

We shall address this problem in three  steps:
\begin{description}
\item{1]} First, we shall discuss the CFT
that describes the Ising model at the critical point. We shall 
 list all operators that may appear 
 as irrelevant ones in the lattice Ising model.
\item{2]}
Then, we shall compare the CFT predictions with the exact results for the
free energy and for the magnetization and with the results for the 
susceptibility reported in~\cite{g2000}. 
We shall see that these results are in perfect agreement 
with the RG and CFT, but have also peculiar 
features that can be explained if we make some additional hypotheses.
The existence in the nearest-neighbor Ising model of exact transformations 
that map the high-temperature phase onto the low-temperature one
(duality or inversion transformations) plays here a major role,
indicating that these peculiar features are strictly 
related to the (partial) solubility of the model.

\item{3]} The conclusions reached in the analysis of the infinite-volume
free energy and of its derivatives are further strengthened by the analysis 
of the mass gap (exponential correlation length) and of the finite-size scaling
of the free energy and of its thermal derivatives at the critical 
point (we use here the results of \cite{ih-00-strip,ih-00-square,Salas-01}).
Finally, we analyze the finite-size scaling of the susceptibility at 
the critical point, showing that the dependence on the boundary
conditions is in perfect agreement with the conjectures we have made.
\end{description}

Since the analysis is rather involved and the reader could be lost in the
technical details of the forthcoming sections, we anticipate here our main
findings:
\begin{itemize}
\item We do not find any evidence for the presence of the leading spin-zero
irrelevant operator predicted by CFT, the energy-momentum tensor. This 
result was already anticipated in \cite{CHPV-gstar,CCCPV-00,CHPV-eqst}
for the
two-dimensional square-lattice Ising model and 
in \cite{r87a} for the one-dimensional Ising quantum chain.
Also, on the triangular lattice we do not observe the next-to-leading 
spin-zero irrelevant operator that has RG dimension $y=-6$.
\item
As mentioned above,
we find unambiguous evidence of the presence of 
nonzero-spin irrelevant operators in the spectrum.
This is not surprising, since such operators are those
that describe the lattice breaking of the rotational symmetry.
What is surprising is that all 
results can be explained in terms of the following conjecture:

{\sl ``The only irrelevant operators which appear in the 
two-dimensional nearest-neighbor Ising model
are those due to the lattice breaking of the rotational symmetry."}

In some sense it can be considered as a renewed version of the original idea of 
Aharony and Fisher.

\end{itemize}

Note that this conjecture applies only to the Ising model with 
nearest-neighbor interactions and it is not known whether 
other formulations of the Ising model satisfy the same conjecture 
(probably they don't!). Moreover, one must in principle distinguish 
between different lattice types. We find that both the square-lattice and 
the triangular-lattice results are compatible with the conjecture, but 
it remains to be understood if it may also hold on other less 
canonical lattices, for instance for honeycomb or Kagom\'e lattices.

This paper is organized as follows. In Sec.~\ref{sec2} we describe the model, 
set our notations, and report the basic results that are needed in the 
following analysis. In Sec.~\ref{sec3} 
we report the CFT analysis of  the 
model at criticality and classify the possible irrelevant operators.
In Sec.~\ref{sec4} we discuss the infinite-volume free energy and its derivatives 
with respect to $h$ for $h=0$. We show that the exact results and the 
results of \cite{g2000} have properties that cannot be anticipated from 
CFT and RG alone. In order to explain them, we put forward four conjectures
that are justified in Sec. \ref{sec4.2} on the basis of the available results.
In Sec. \ref{sec4.3}, on the basis of the conjectures we have made, we obtain 
some general predictions for the susceptibility on the triangular lattice. 
The extension of the results of \cite{g2000} to such a lattice
is very important in order to understand the validity of our conjectures.
In Sec. \ref{sec5} we discuss the critical behavior of the 
exponential correlation length. The analysis on the triangular lattice 
is particularly interesting and gives strong support to the conjecture 
we have presented above. 
In Sec. \ref{sec6} and \ref{sec7} we consider the finite-size scaling 
of several quantities at the critical point. We show that the existence of an 
inversion (duality) transformation and the general conjecture 
presented above explain some peculiar features of the results 
found in \cite{Salas-01,ih-00-strip,ih-00-square}. 
In Sec. \ref{sec8} we summarize the results 
and discuss some open problems.

\section{The Ising model with nearest-neighbor interactions}
\label{sec2}

The two-dimensional Ising model is defined by the partition function
\eq
Z=\sum_{\sigma_i=\pm1}e^{\beta\sum_{\br n,m \kt}\sigma_n\sigma_m
+h\sum_n\sigma_n} ,
\label{zz1}
\en
where the spin variables $\sigma_n$ are defined on the sites $n$ of a regular
lattice and take the values $\{\pm 1\}$.
The model has two phases: the low-temperature one, in which the 
$\sZ_2$ symmetry is spontaneously broken and the high-temperature one in 
which the symmetry is restored. The two phases are separated by a critical 
point which is located at $\beta=\beta_c$. 

In the following we will study several observables. We define\footnote{
Note that our definitions differ by powers of the temperature 
and by signs from the usual thermodynamic ones. This is irrelevant for 
our purposes.} the 
free-energy density $F(\beta,h)$, the energy per site $E(\beta,h)$,
the specific heat $C(\beta,h)$, the magnetization per site $M(\beta,h)$,
and the susceptibility $\chi(\beta,h)$:
\begin{eqnarray}
F(\beta,h) &\equiv& \lim_{N\to\infty} {1\over N} \log(Z(\beta,h)), 
\label{F-definition} \\
E(\beta,h) &\equiv& -  {\partial F(\beta,h)\over \partial \beta}, \\ 
C(\beta,h) &\equiv& {\partial^2 F(\beta,h)\over \partial \beta^2}, \\ 
M(\beta,h) &\equiv& {\partial F(\beta,h)\over \partial h}, \\
\chi(\beta,h) &\equiv& {\partial^2 F(\beta,h)\over \partial h^2}. 
\end{eqnarray}
In \reff{F-definition} $N$ is the number of sites of a finite lattice.

\subsection{The square lattice} \label{sec2.1}

On the square lattice 
\be
\beta_c=\smfrac12\log{(\sqrt{2}+1)}=0.4406868\ldots
\ee
and we will measure the deviations from the critical temperature in 
terms of the variable $\tau$ introduced in \cite{g2000}:
\be
\tau = \frac12 \left({1\over \sinh 2\beta} - \sinh 2\beta \right).
\label{tau-square}
\ee 
For $\beta=\beta_c$, $\tau =0$, while $\tau > 0$ (resp. $\tau < 0$) for 
$\beta < \beta_c$ (resp. $\beta > \beta_c$).

We will use the exact expressions for the free-energy density and magnetization 
in zero field given by \cite{mccoy}
\begin{eqnarray}
F(\tau,0) &=& \smfrac12 \log\left(2 \cosh^2 2\beta\right) + 
         F^{\rm sing}(\tau), \\
M(\tau,0) &=& \left(1 - k(\tau)^2\right)^{1/8},
\label{magnetization-h0}
\end{eqnarray}
where
\begin{eqnarray}
F^{\rm sing}(\tau) &=& \int_0^\pi {d\theta\over 2\pi}\, 
\log\left[1 + \left(1 - {\cos^2\theta\over 1 + \tau^2}\right)^{1/2}\right],
\\
k(\tau) &=& \left(\sqrt{1 + \tau^2} + \tau\right)^2.
\end{eqnarray}
In this work, the duality transformation that maps the high-temperature phase 
onto the low-temperature one plays an important role. 
The variable $\tau$ transforms naturally under such transformation, 
i.e. $\tau \to -\tau$. It is easy to verify that 
\begin{eqnarray}
  k(-\tau) &=& {1\over k(\tau)},
\label{k-duality} \\
F^{\rm sing}(-\tau) &=& F^{\rm sing}(\tau) , 
\label{Fsing-duality}\\
 k(-\tau)^{-1/8} (-\tau)^{-1/8} M(-\tau,0) &=& 
 k(\tau)^{-1/8} \tau^{-1/8} M(\tau,0).
\label{magnetization-duality}
\end{eqnarray}
By using the exact expressions for the free energy and the magnetization
we define two functions $a(\tau)$ and $b(\tau)$ that will play 
a major role below. They are defined by requiring
\begin{eqnarray}
F(\tau,0) &=& - A a(\tau)^2 \log |a(\tau)| + A_0(\tau),
\label{def-function-a} \\
M(\tau,0) &=& B b(\tau) |a(\tau)|^{1/8},
\label{def-function-b}
\end{eqnarray}
where $a(\tau)$, $b(\tau)$, and $A_0(\tau)$ are regular functions\footnote{
We will call a function
{\em regular} if it has an expansion in integer powers of $\tau$ for
$\tau\to0$.} of $\tau$,
$a(\tau)\approx \tau$ for $\tau\to 0$, $b(0) = 1$, and
$A$ and $B$ are constants. 
Explicitly we find
\begin{eqnarray}
a(\tau) &=& \tau \left( 1 - {3\over16} \tau^2 + {137\over 1536} \tau^4 +
        O(\tau^6) \right), 
\label{a-function-square}\\
b(\tau) &=& k(\tau)^{1/8} \left(1 + {11\over 128}  \tau^2
               - {3589\over 98304 } \tau^4 + O(\tau^6)\right),
\end{eqnarray}
and
\be
A = {1\over 2\pi}, \qquad\qquad B = 2^{1/4}.
\ee 
Under duality,
\be
a(-\tau)= -a(\tau) \qquad\qquad
k(-\tau)^{-1/8} b(-\tau) = k(\tau)^{-1/8} b(\tau).
\label{duality-a-b}
\ee                                                                             
Although the susceptibility in zero field has not been computed exactly,
its behavior for $h=0$, $\tau \to 0$ is quite well known. In  
\cite{g2000} the asymptotic behavior of $\chi$ for $h=0$ in both phases
was obtained:
\be
\chi_{\pm}(\tau) = C^{\pm} |\tau|^{-7/4} k(\tau)^{1/4}
     \widehat{F}_\pm(\tau) + B_f(\tau),
\label{Orrick-chi}
\ee
where $\widehat{F}_\pm(\tau)$ are regular
functions of $\tau$,
\begin{eqnarray}
B_f(\tau) &=& \sum_{q=0}^\infty \sum_{p=0}^{\lfloor \sqrt{q}\rfloor}
            b^{(p,q)} \tau^q (\log |\tau|)^p,
\label{Bf-def}
\end{eqnarray}
and $\tau$ is defined in \reff{tau-square}. Here $\chi_+(\tau)$ 
($\chi_-(\tau)$) is 
the susceptibility in the high- (low-) temperature phase.
 
By a careful numerical study, reference \cite{g2000} found two additional
important properties of $\widehat{F}_\pm(\tau)$.
First, $\widehat{F}_\pm(\tau)$ are even functions of $\tau$. There is no
rigorous proof, but we note that a similar
property is satisfied by the two-point function in the large-$x$
limit, see Sec. \ref{sec5.1}.                                                                         
Moreover, the results of \cite{g2000} can be written as
\be
\widehat{F}_\pm (\tau) = \left[ a(\tau) \tau^{-1}\right]^{-7/4}
           \left[b(\tau) k(\tau)^{-1/8}\right]^2 G_\pm (a(\tau)),
\label{hatF-square}
\ee
where $G_\pm (z) $ are even functions of $z$, and $a(\tau)$
and $b(\tau)$ are defined in Eqs. \reff{def-function-a},
\reff{def-function-b}. Explicitly
\be
G_\pm (z) = 1 - {1\over 384} z^4 + \left(f_\pm^{(6)} - {49\over 1536}\right)
          z^6 + O(z^8),
\label{G-square}
\ee
where $f_\pm^{(6)}$ are numerical coefficients reported in
\cite{g2000}.
Note the absence of the term of order $z^2$, a result that will play
a major role below.

\subsection{The triangular lattice} \label{sec2.2}

On the triangular lattice 
\be
\beta_c = \smfrac14 \log 3 = 0.2746531\ldots
\ee
We measure the deviations from the critical temperature in terms of the 
variable $\tau$ defined by 
\be
\tau \equiv {1 - 4 v + v^2\over \sqrt{2 v} (1 - v)},
\label{tau-tria}
\ee
where $v \equiv \tanh \beta$.
Under the inversion
transformation that maps the high-temperature phase onto the
low-temperature one,
\be
v \to v' = \left({\sqrt{1 - v + v^2} - \sqrt{v}\over (1 - v)}\right)^2,
\ee
it transforms simply as $\tau \to - \tau$. It is thus the
analogous of the variable \reff{tau-square} introduced in \cite{g2000}. 

In zero field, the free-energy density is given by \cite{Stephenson-64}
\be
F(\tau,0) = \smfrac12 \log(4 \sinh 2\beta) + F^{\rm sing}(\tau),
\ee
where 
\be
F^{\rm sing}(\tau) = 
    {1\over2} \int_0^{2\pi}\int_0^{2\pi} {d\phi_1\over 2\pi}\, 
      {d\phi_2\over 2\pi}\, 
 \log\left[3 + \tau^2 - \cos\phi_1 - \cos\phi_2 - \cos(\phi_1 + \phi_2)\right],
\ee
the magnetization by \reff{magnetization-h0}, where \cite{Stephenson-64}
\be
k(\tau) =  {(1 - v)^3 (1 + v)\over 4  v \sqrt{v (1 - v + v^2)}}.
\ee                                                                             
Under $\tau\to-\tau$, relations \reff{k-duality}, \reff{Fsing-duality}, 
and \reff{magnetization-duality} hold on the triangular lattice too.

{} From the expressions of the magnetization and of the free energy,
we can compute the functions $a(\tau)$ and $b(\tau)$ that are defined by
\reff{def-function-a} and \reff{def-function-b}. In this case
we obtain
\begin{eqnarray}
\hskip -1truecm
a(\tau) &=& \tau - {\tau^3\over 24} + {47 \tau^5\over 10368} -
         {161 \tau^7\over 248832} + {113191 \tau^9\over 1074954240} +
        O(\tau^{11}), 
\label{a-function-tria} \\
\hskip -1truecm
b(\tau) &=& k(\tau)^{1/8}
   \left(1 + {11\tau^2\over288} - {671\tau^4\over165888} +
             {10115\tau^6\over15925248} -
             {31791497\tau^8\over275188285440} + O(\tau^{10})\right),
\end{eqnarray}                                                                  and
\be
A = {1\over 2 \sqrt{3} \pi}, \qquad\qquad
B = \left({8\over3}\right)^{1/8}.
\ee
As in the square-lattice case, the functions
$a(\tau)$ and $b(\tau)$  satisfy the duality relations
\reff{duality-a-b}.

\section{Conformal field theory analysis}
\label{sec3}

\subsection{Primary and secondary fields} \label{sec3.1}

The Ising model at the critical point is described 
 by the unitary minimal CFT with central charge
$c=1/2$~\cite{bpz}.
Its spectrum  can be divided into three conformal families characterized by
different transformation 
properties under the dual and $\sZ_2$ symmetries of the model. They
 are the identity, spin, and energy families and are
 commonly denoted as  $[{I}],~[\sigma],~[\epsilon]$.
Let us discuss their features in detail.
\begin{itemize}
\item{\bf Primary fields}

Each family contains an 
operator which is called primary field (and gives the name to the entire
family).  
 Their conformal weights are $h_{I}=0$,
 $h_\sigma=1/16$ and  $h_\epsilon=1/2$ respectively. 
 Since the RG eigenvalues are related to the conformal weights by $y=2-2h$,
 all primary fields are relevant.

\item{\bf Secondary fields}

All the remaining operators of the three
 families (which are called secondary fields) are generated from the primary
 ones by applying the generators $L_{-i}$ and $\bar L_{-i}$ 
of the Virasoro algebra defined by
\eq
[L_n,L_m]=(n-m) L_{n+m} +\frac{c}{12} n(n^2-1)\delta_{n+m,0}\; .
\label{vir}
\en

It can be shown that, by applying  
a generator of index $k$, $L_{-k}$ or $\bar L_{-k}$, to a field $\phi$ 
(where $\phi={I},\epsilon,\sigma$ depending on the case)
of conformal weight $h_\phi$, a  new operator of weight
$h=h_\phi+k$ is obtained.
In general, any combination of $L_{-i}$ and $\bar L_{-i}$
is allowed.
If we denote with $n$ the sum of the indices of the generators of 
type $L_{-i}$
and with $\bar n$ the sum of those of type $\bar L_{-i}$, the conformal 
weight of the resulting operator is $h_\phi+n+\bar n$.
The corresponding RG eigenvalue is $y=2-2h_\phi-n-\bar n$.

\item{\bf Nonzero spin states}

The secondary fields may have nonzero
spin, which is given by the difference  $n-\bar n$. In general, one is
interested in quantities that are invariant under the lattice rotation
group, and thus in operators that belong to its identity representation.
Since the lattice invariance group is a finite 
subgroup of the rotation group, in the lattice discretization of 
a scalar operator, operators that do not have spin zero, 
i.e. transform nontrivially for general rotations, may appear.
The invariance group of the square lattice 
is the finite subgroup $C_4$ (cyclic group of order four),
which has four representations of ``discrete" spin 0, 1, 2, and 3. 
An observable that transforms as a spin-$j$ representation under the 
full rotation group belongs to a representation of discrete spin
$j~({\rm mod}~4)$ under the action of $C_4$. Therefore, a lattice 
scalar operator is expressed as a sum of continuum operators 
of spin $4 j$, $j\in {\sN}$. 
Analogously, on a triangular lattice the rotation group is broken to 
the cyclic group of order six $C_6$. In this case, a lattice scalar 
operator is expressed in terms  of continuum operators
of spin $6 j$, $j\in {\sN}$.

\item{\bf Null vectors}

 Some of the secondary fields disappear from the spectrum due to the null-vector
 conditions (see~\cite{bpz}).
 In particular, this happens for one of the two states at level 2 in
 the $[\sigma]$ and $[\epsilon]$ families and for the unique state at level 1 in
 the identity family. From each null state one can generate, by applying the
 Virasoro operators, a whole family of null states. Hence, at level 2 in the
 identity family there is only one surviving secondary field, which can be
 identified with the stress-energy tensor $T$ (or $\bar T$). The second null
 vector in the $\sigma$ family appears at level 3 while in the $\epsilon$
 family it appears at level 4. This fact will play an important role in the
 following.

\item{\bf Secondary fields generated by $L_{-1}$}

 Among all secondary fields, a particular role is played by those generated
 by the $L_{-1}$ Virasoro generator. $L_{-1}$ is the generator of 
 translations on the lattice and as a consequence, it has zero eigenvalue on
 translationally invariant observables. Another way to state this result is 
 that $L_{-1}$ can be represented
 as a total derivative, and as such it gives zero if applied to an operator
 which is the integral of a suitable
 density over the lattice, i.e. a translationally invariant operator. 

\item{\bf Quasiprimary operators.}

A quasiprimary field $|Q>$ is a secondary field which satisfies the equation
\eq
L_1|Q>=0\; .
\label{t1}
\en
This condition eliminates all the  secondary fields which are
generated by $L_{-1}$. The quasiprimary operators are the only ones that may 
appear in translationally invariant quantities.


\end{itemize}

\subsection{Quasiprimary states and irrelevant operators.} \label{sec3.2}

It is easy to construct, 
by using (\ref{t1}), all the low-lying quasiprimary states. 
Here is the list of all quasiprimary operators up to level 10. 
\begin{itemize}
\item
In the Identity family there is one quasiprimary state at levels 2, 4, and 6
and two quasiprimary states at levels 8 and 10;
\item
In the energy family there is one quasiprimary state at levels 4, 6, 7, 8, and 9
and two quasiprimary states at level 10;
\item
In the $[\sigma]$ family there is one quasiprimary state at levels 3, 5, 6, 7,
and 8 and two quasiprimary states at levels 9 and 10.
\end{itemize}
For all these states it is possible to give the exact expression in terms of
the Virasoro generators (even if it becomes increasingly cumbersome as the 
level increases). 
For instance, in the identity family one finds
\eq
Q_2^{I}=L_{-2}|{I}>,
\en
\eq
Q_4^{I}= \left(L_{-2}^2-\smfrac35 L_{-4}\right)|{I}>,
\en
at level 2 and 4 respectively,
where we have introduced the
 notation $Q^{\eta}_n$ to denote the quasiprimary state at
level $n$ in the $\eta$ family.

Let us now construct from the $Q^{\eta}_n$ listed above the irrelevant operators
which could appear in any lattice translationally invariant quantity. 
We list below those that have RG eigenvalue $|y| < 10$.
We will classify them by their spin,
since operators of different spin appear on different lattices. 
Spin-zero operators are relevant in all cases, spin-$(4n)$
operators appear on the square lattice, while spin-$(6n)$ operators 
play a role only on the triangular lattice.

The spin-0 operators are the following:
\begin{itemize}
\item Identity family:
  
 $Q_2^{I} \bar{Q_2^{I}}$ whose weight is 4 and RG eigenvalue is $-2$;

 $Q_4^{I} \bar{Q_4^{I}}$ whose weight is 8 and RG eigenvalue is $-6$;


\item Energy family:

 $Q_4^{\epsilon} \bar{Q_4^{\epsilon}}$ whose weight is 9 and RG eigenvalue is 
  $-7$;

\item Spin family:

 $Q_3^{\sigma} \bar{Q_3^{\sigma}}$ whose weight is $6+\frac18$
 and RG eigenvalue is $-(4+\frac18)$;

 $Q_5^{\sigma} \bar{Q_5^{\sigma}}$ whose weight is $10+\frac18$
 and RG eigenvalue is $-(8+\frac18)$.

\end{itemize}

On the square lattice we should consider the spin-four operators:
 \begin{itemize}

\item Identity family:
  
 $Q_4^{I} + \bar{Q_4^{I}}$ whose weight is 4 and RG eigenvalue is $-2$;

 $Q_6^{I} \bar{Q_2^I} + Q_2^{I} \bar{Q_6^{I}}$ 
 whose weight is 8 and RG eigenvalue is $-6$;

\item Energy family: 

 $Q_4^{\epsilon} + \bar{Q_4^{\epsilon}}$ 
  whose weight is 5 and RG eigenvalue is $-3$.

\item Spin family:
 $\bar Q^\sigma_3 Q^\sigma_7 + \bar Q^\sigma_7 Q^\sigma_3$
  whose weight is $10+\frac18$ and RG eigenvalue is $-(8+\frac18)$.

\end{itemize}

Also the spin-eight contribute on the square lattice at this order:

 \begin{itemize}

\item Identity family:
  
 $Q_8^{I} + \bar{Q_8^{I}}$ whose weight is 8 and RG eigenvalue is $-6$;

\item Energy family:
  
 $Q_8^{\epsilon} + \bar{Q_8^{\epsilon}}$ whose weight is 9 
  and RG eigenvalue is $-7$;

\item Spin family:
  
 $Q_8^{\sigma} + \bar{Q_8^{\sigma}}$ whose weight is $8+\frac18$
 and RG eigenvalue is $-(6+\frac18)$.

\end{itemize}

On the triangular lattice we should consider the spin-six operators:

\begin{itemize}

\item Identity family:

$Q^{I}_6 + \bar{Q}^{I}_6$ whose weight is $6$
  and RG eigenvalue is $-4$;

$\bar Q^{I}_2 Q^{I}_8 + \bar Q^{I}_8 Q^{I}_2$ whose weight 
is $10$ and RG eigenvalue is $-8$;
  
\item Energy family:

$ Q^\epsilon_6 + \bar Q^\epsilon_6$ whose weight is $7$ 
  and RG eigenvalue is $-5$;
  
\item Spin family:

$Q^\sigma_6 + \bar Q^\sigma_6$ whose weight is $6+\frac18$
  and RG eigenvalue is $-(4+\frac18)$.
  
\end{itemize}

Higher-order spins contribute operators with $y \le -10$. For instance, in the 
identity family one should consider 
the spin-12 operator 
$Q^{I}_{12} + \bar Q^{I}_{12}$ whose weight is $12$
 and RG eigenvalue is $-10$.

Among these operators, the most important ones are: 
$Q_2^{I} \bar{Q_2^{I}}$ that has spin zero and $y=-2$ and should be considered
both for the square and the triangular lattice; 
$Q_4^{I} + \bar{Q_4^{I}}$ (with $y=-2$) and 
$Q_6^{I} + \bar{Q_6^{I}}$ (with $y=-4$) that are the leading operators 
that break rotational invariance on the square and on the triangular lattice
respectively. These operators can 
be explicitly related to the energy-momentum tensor. The relations are:
$Q_2^{I} \bar{Q_2^{I}}= T\bar T$,
$Q_4^{I} + \bar{Q_4^{I}}=T^2+\bar T^2$,
$Q_6^{I} + \bar{Q_6^{I}}=T^3+\bar T^3$.
These operators will play an important role in the following discussion. 

As a general remark, it is important to notice that, since only even-spin 
operators are of interest, the dimensions $y$ of the operators satisfy the 
following conditions: $y\in 2 {\sZ}$ for the identity family,
$y\in 2 {\sZ} + 1$ for the energy family, and 
$y\in 2 {\sZ} - \frac18$ for the spin family.

Finally, we want to discuss the role of the symmetries. On the lattice 
there are two exact symmetries that will play an important role. 
\begin{itemize}
\item $\sZ_2$ symmetry: $(h\to -h)$. Under this transformation the 
operators belonging to the identity and to the energy family are even, while 
the operators belonging to the spin family are odd. 
\item duality (inversion) symmetry for $h=0$. This transformation maps the 
high-temperature phase onto the low-temperature one and with our choice of 
variable $\tau$ (see \reff{tau-square} and \reff{tau-tria} for the 
square and the triangular lattice respectively)
it corresponds to the mapping $\tau\to -\tau$. Under this transformation 
(see, e.g., Appendix E of \cite{bpz})
the identity operators are even, the energy operators are odd, while the 
$[\sigma]$-family operators do not have a well-defined behavior.
\end{itemize}

\section{Infinite-volume zero-momentum quantities for\\ $h=0$}
\label{sec4}

In this  Section, using the results of 
Sec. \ref{sec3}, we shall derive the scaling behavior of the 
free energy, magnetization, and susceptibility at $h=0$ and we will
compare these results with the exact expressions for 
$F(\tau,0)$ and $M(\tau,0)$ and with the results of \cite{g2000}
on the square lattice. We will verify that the structure of these 
expressions is in agreement with the RG
predictions, although the 
complicated logarithmic dependence found in \cite{g2000} 
requires an extension of the usual scaling expressions. 
Moreover, the exact results and those of \cite{g2000}
have additional properties that are 
specific of the lattice nearest-neighbor Ising model and 
are probably not satisfied 
by a generic model belonging to the Ising universality class. 
All these properties can be explained if we make some general 
conjectures: they will be presented in Sec. \ref{sec4.1}.

We present a general analysis for the square and the triangular lattice.
In particular, we will show that the extension of the work of 
\cite{g2000} to the triangular lattice would provide
strong support for (or rule out) our conjectures.

\subsection{Renormalization-group predictions and conjectures} 
\label{sec4.1}

We wish now to derive the asymptotic behavior of $F(\tau,0)$, $M(\tau,0)$,
and $\chi(\tau,0)$ by using the RG approach
and the classification of the irrelevant 
operators presented in Sec. \ref{sec3.2}. 
We write the free energy as \cite{Wegner-76}
\begin{eqnarray}
F(\tau,h) & = & f_{b}(\tau,h) + 
   |u_t|^{2/y_t} f_{\pm}\left(\left\{\frac{u_j}{|u_t|^{y_j/y_t}}\right\}\right)
\nonumber \\
&&  +
   |u_t|^{2/y_t} \log |u_t| 
   \widetilde{f}_{\pm}\left(\left\{\frac{u_j}{|u_t|^{y_j/y_t}}\right\}\right),
\label{F-RG}
\end{eqnarray}
where $f_{b}(\tau,h)$ is a regular function\footnote{Sometimes
it is assumed that the bulk free energy depends on the temperature 
only \cite{Privman-90,PHA-91}. However, this conjecture is 
inconsistent with the rigorous results available for $\chi$. 
See \cite{ssv1} for a critical discussion.} of $\tau$ and $h^2$, 
$u_t$ and $u_j$ are nonlinear scaling fields associated with the temperature
and with all other operators with corresponding dimensions $y_t=1$ and 
$y_j$. They include the nonlinear scaling field associated with the 
magnetic field with dimension $y_h = 15/8$ and those associated with  
all irrelevant operators. Note the presence of the logarithmic term
due to a resonance\footnote{Since secondary fields belonging to a given family
differ by integers, we expect additional multiple resonances and 
additional terms with higher powers of $\log |u_t|$ in Eq. \reff{F-RG}.
Such higher powers have indeed been found in the analysis of $\chi$ 
\cite{g2000}.}
 between the thermal and the identity operator
which is responsible of the log-type singularity in the specific heat
\cite{Wegner-76}.
The nonlinear scaling fields are analytic functions of $\tau$ and $h$
that have well-defined transformation properties under $h\to -h$. 
Those associated with the identity and the energy family 
are even under the transformation, while those associated with the $[\sigma]$
family (and thus $u_h$ too) are odd. For our purposes we can expand
\begin{eqnarray}
   u_t(\tau,h) &=& \mu_t(\tau) + {h^2\over2} \lambda_t(\tau) + O(h^4), \\
   u_j^{\rm even}(\tau,h) &=& \mu_j(\tau) + {h^2\over2} \lambda_j(\tau) + 
          O(h^4), \\
   u_j^{\sigma}(\tau,h) &=& h v_j(\tau) + O(h^3).
\label{def-vh}
\end{eqnarray}
The ${\sZ}_2$-even operators belong to the identity and the 
energy family and thus,
for $h=0$, they have well-defined properties under duality:
\begin{eqnarray}
   \mu_t(-\tau) &=& - \mu_t(\tau), \nonumber \\ 
   \mu_j^{\epsilon}(-\tau) &=& - \mu_j^{\epsilon}(\tau), \nonumber \\ 
   \mu_j^{I}(-\tau) &=&  \mu_j^{I}(\tau). 
\label{scaling-fields-duality}
\end{eqnarray}
In general, we expect $\mu_j^{I}(0)\not=0 $, 
and therefore we can normalize these scaling fields by requiring
$\mu_j^{I}(0)=1$. On the other hand, the energy-family 
scaling fields---including that associated with the temperature---vanish for 
$\tau=0$ and thus we normalize them by requiring 
$\mu_j^{\epsilon}(\tau)\approx\tau$. The spin-family fields are normalized 
by requiring $v_j(0)=1$.

Let us now present our basic conjectures that will be justified in 
Sec. \ref{sec4.2} on the basis of the exact expressions for the 
free energy and the magnetization and of the results of \cite{g2000}. 
Two conjectures will be presented in different forms.
The analysis reported here of the 
infinite-volume quantities gives only evidence for the weaker versions
(c1) and (d0). 
Evidence for (c2) will be provided in Sec. 6, and evidence for (d1)/(d2) 
in Sec. \ref{sec5.2}. As we will discuss, the analysis of $\chi$ on the 
triangular lattice should be able to discriminate between (d1) and (d2).

Let us now give the list of the conjectures:
\begin{itemize}
\item[(a)] Consider a $[\sigma]$-family operator, and let 
$v_j(\tau)$ be the corresponding nonlinear scaling field for
$h \to 0$, cf. \reff{def-vh}. Then,
either $v_j(\tau)=0$, i.e.
the corresponding operator is decoupled, or
\be
 k(-\tau)^{-1/8} v_j(-\tau) = k(\tau)^{-1/8} v_j(\tau).
\label{property-vj}
\ee
Such a relation
should be satisfied by $v_h(\tau)$ since the corresponding 
 operator does not decouple. 
\item[(b)]
The functions $f_\pm$ and $\tilde{f}_\pm$ are even functions of
the nonlinear scaling fields associated with the energy family.
\item[(c1)] The functions $\tilde{f}_\pm$ depend only on the $\sZ$-even scaling
fields.
\item[(c2)] Stronger version of the previous one: The functions 
$\tilde{f}_\pm$ are constant. Such a conjecture was already made by 
Aharony and Fisher \cite{af}.
\item[(d0)] The nonlinear scaling field of the $T\bar{T}$ operator
vanishes at the critical point: $u_{T\bar{T}}(0,0) = 0$.
\item[(d1)] Stronger version of (d0): The operator $T\bar{T}$ decouples,
i.e. $u_{T\bar{T}}(\tau,h) = 0$ for all $\tau$ and $h$.
\item[(d2)] Stronger version of (d1): The only irrelevant operators that 
appear in the Ising model are the non-rotationally invariant ones.
\end{itemize}
We remark that these conjectures (in their stronger form) 
are sufficient to explain the existing data, but are by no means necessary. 
For instance, consider the three conjectures (d).
All existing square-lattice results require only (d0). 
Conjectures (d1) and (d2) are supported by the results on the 
triangular lattice that will be presented in Sec. \ref{sec5.2} and 
\ref{sec6}. There we will show $\mu_{T\bar{T}}(\tau) = o(\tau^4)$, 
which provides evidence for (d1), and $\mu(0) = 0$ 
for the scalar operator $Q_4^I \bar{Q}_4^I$  with $y=-6$, 
which is our motivation for the conjecture (d2). 
We wish also to stress that, at least in principle, some properties
may hold only on a very specific lattice type and thus the 
observed properties on the triangular lattice may not 
extend to the square-lattice case.

Let us note that in the analysis of the scaling corrections the spin
of the operator will play an important role. As we already mentioned 
in Sec. \ref{sec3.1}, all operators of spin $4j$ (respectively $6j$) 
appear in \reff{F-RG} on the square (resp. triangular) lattice, 
$j\in \sN$. However, because of the rotational invariance of the 
critical theory, nonzero spin operators contribute only at second 
order in the Taylor expansion of the infinite-volume free energy in powers 
of $u_j |u_t|^{-y_j/y_t}$. 

\subsection{The square lattice} \label{sec4.2}

Let us now use the exact results for $F(\tau,0)$ and $M(\tau,0)$ and 
the results of \cite{g2000} to provide evidence for the 
conjectures we made in the previous section.

Setting $h=0$ in \reff{F-RG} we see that all scaling fields associated 
with the $[\sigma]$ family disappear. 
Since the dimensions of the operators belonging to the energy and to the 
identity family are integers we predict
\be
F(\tau,h=0)_\pm = f_0(\tau) + f_1(\tau) \log|\tau|,
\ee
where $f_0(\tau)$ and $f_1(\tau)$ have a regular expansion in $\tau$.
The functions $f_0(\tau)$ and $f_1(\tau)$ can in principle depend on the phase,
but from the exact solution we know that this is not the case. This implies
\begin{eqnarray}
\phi(\{x_j\}) &\equiv& f_+\left(\{x_j\}^{{I},\epsilon}; \{x_j=0\}^{\sigma} \right) 
          = f_-\left(\{x_j\}^{{I},\epsilon}; \{x_j=0\}^{\sigma} \right), 
\label{def-phi} \\
\widetilde{\phi}(\{x_j\}) &\equiv& 
      \widetilde{f}_+\left(\{x_j\}^{{I},\epsilon}; \{x_j=0\}^{\sigma} \right) 
    = \widetilde{f}_-\left(\{x_j\}^{{I},\epsilon}; \{x_j=0\}^{\sigma} \right).
\label{def-phitilde}
\end{eqnarray}
Using \reff{Fsing-duality}, we find that $f_1(\tau)$ is even in $\tau$, 
a property that is 
certainly satisfied if the conjecture (b) is true, i.e. 
$\widetilde{\phi}(\{x_j\})$ is an even function 
of the energy-family scaling fields. If this is true, the energy-family 
scaling fields would begin to contribute to second order.

Let us now consider the magnetization in the low-temperature phase.
{}From  \reff{F-RG} we obtain ($\tau < 0$)
\be
M(\tau) = \sum_{k\,\in [\sigma]} 
        |\mu_t|^{2-y_k} v_k \rho_k(\{\mu_j \mu_t^{-y_j}\}^{{I},\epsilon}) + 
    \log|\mu_t|\ \sum_{k\,\in [\sigma]}
     |\mu_t|^{2-y_k} v_k \tilde{\rho}_k(\{\mu_j \mu_t^{-y_j}\}^{{I},\epsilon}),
\ee
where the functions $\rho_k$ and $\tilde{\rho}_k$
 depend only on the scaling fields of the 
${\sZ}_2$-even operators, and the sums are over all $[\sigma]$-family operators. 
Now, if $y_k$ is the dimension of an operator 
belonging to the $[\sigma]$ family, $y_k = -1/8 + 2n$, 
where $n$ is an integer.
Therefore, we predict
\be
M(\tau) = (-\tau)^{1/8} M_0(\tau) + (-\tau)^{1/8} M_1(\tau) \log (-\tau),
\ee
where $M_0(\tau)$  and $M_1(\tau)$ are regular functions of $\tau$. 
Now, the exact solution gives $M_1(\tau) = 0$, a property that is 
satisfied if the conjecture (c1) is true.
Setting $M_1(\tau) = 0$, we find a perfect agreement
with the exact result.

However, the exact result satisfies an additional property:
Using \reff{magnetization-duality}, we have
\be
  k(-\tau)^{-1/8}  M_0(-\tau) = M_0(\tau) k(\tau)^{-1/8}.
\ee
By using the fact that $y_j = 2n-\frac18$ 
(resp. $y_j=2n-1$, $y_j=2n$) for a $[\sigma]$ (resp. $[\epsilon]$, $[I]$) 
family
operator, $n\in \sZ$, it is easy to verify that such an equation 
is automatically satisfied if the conjectures (a) and (b) are true. 

Let us consider the susceptibility. By differentiating \reff{F-RG} 
and using Eqs. \reff{def-phi} and \reff{def-phitilde}, we obtain
\begin{eqnarray}
\hskip -1truecm
\chi_\pm &= & \left. {\partial^2 f_b\over \partial h^2}\right|_{h=0}  + 
      \mu_t \lambda_t \left[2 \phi(\{x_j\}) + \tilde{\phi}(\{x_j\}) \right]
    + \mu_t^2 \sum_{ik\,\in [\sigma]} \psi_{ik,\pm}(\{x_j\}) v_i v_k 
           |\mu_t|^{-y_i-y_k} \nonumber \\
&&  + \mu_t^2 \sum_{k\,\in [I],[\epsilon]} {\partial \phi\over \partial x_k} 
                   (\{x_j\}) |\mu_t|^{-y_k} 
           \left(\lambda_k - y_k \mu_k \lambda_t \mu_t^{-1}\right) + 
       2 \mu_t \lambda_t \tilde{\phi}(\{x_j\}) \log |\mu_t| 
\nonumber \\
&& 
    + \mu_t^2 \log |\mu_t|
       \sum_{ik\,\in [\sigma]} \tilde{\psi}_{ik,\pm}(\{x_j\}) v_i v_k 
           |\mu_t|^{-y_i-y_k}  
\nonumber \\
&&    + \mu_t^2 \log |\mu_t|\sum_{k\,\in [I],[\epsilon]} 
             {\partial \tilde{\phi}\over \partial x_k} 
                   (\{x_j\}) |\mu_t|^{-y_k} 
           \left(\lambda_k - y_k \mu_k \lambda_t \mu_t^{-1}\right),
\end{eqnarray}
where all functions depend only on the irrelevant ${\sZ}_2$-even scaling 
fields through $x_j = \mu_j \mu_t^{-y_j}$,
$\phi$ and $\tilde{\phi}$ are defined in 
Eqs. \reff{def-phi}, \reff{def-phitilde}, and 
$\psi_{ik,\pm}$ and $\tilde{\psi}_{ik,\pm}$ are second-order
derivatives of $f_\pm$ and $\tilde{f}_\pm$ with respect to the 
$[\sigma]$-family fields.
The sums over $\sZ_2$-even fields include only the irrelevant 
ones---the temperature 
should be excluded---while the sums over $[\sigma]$-fields include both the 
magnetic and the irrelevant ones.
Since $y_j = -1/8 + 2n$, $n$ integer, for $[\sigma]$ operators and $y_j$ integer for 
${\sZ}_2$-even operators, this result implies the expansion 
\be
\chi_{\pm} = |\tau|^{-7/4} A_\pm(\tau) + 
             |\tau|^{-7/4} \log |\tau| B_\pm(\tau) + C(\tau) +
             D(\tau) \log |\tau|,
\label{RGprediction-chi}
\ee
where all functions are regular and only $A_\pm$ and $B_\pm$ depend on the 
phase.  

If we now use the conjecture (c1) we obtain
$\tilde{\psi}_{ik,\pm} = 0$, and therefore $B_\pm(\tau) = 0$ in agreement 
with the results of \cite{g2000}.

By comparing \reff{RGprediction-chi} with \reff{Orrick-chi}, 
we find $B_f(\tau) = C(\tau) +
             D(\tau) \log |\tau|$,  so that
$B_f(\tau)$ should be identical in both phases,
in agreement with the results of \cite{g2000}.
However, we predict only a single $\log |\tau|$, while in 
\cite{g2000} all powers appear. This means that our scaling Ansatz 
\reff{F-RG} is not correct: There are additional resonances that give rise 
to a more complicated logarithmic structure. 

For $\widehat{F}_\pm(\tau)$ we find
\be
\widehat{F}_\pm(\tau) = 
    {1\over C^\pm} k(\tau)^{-1/4} \tau^4 \left(\mu_t\over\tau\right)^{2+1/4} 
     \sum_{ik\, \rm odd} \psi_{ik,\pm}(\{x_j\}) v_i v_k 
           \mu_t^{-y_i-y_k-1/4}\; .
\ee
By using the conjectures (a) and (b), we can show that 
$\widehat{F}_\pm(\tau)$ is even in $\tau$, in agreement with the 
results of \cite{g2000}. Note that the functions $\lambda_j(\tau)$ 
instead have no specific properties under $\tau\to -\tau$ and indeed 
$B_f(\tau)$ contains all powers of $\tau$. 

Let us now discuss in more detail the consequences of Eqs. 
\reff{hatF-square} and \reff{G-square}. First, notice that the most 
important irrelevant operator of the $[\sigma]$ family 
($Q_3^{\sigma} \bar{Q_3^{\sigma}}$) has dimension $y=-4-1/8$. 
Since $y_h = 2 - 1/8$, it gives corrections of order
$\tau^6$. Thus, neglecting corrections of this order, we need to consider 
only the magnetic operator (the leading one) 
among the $[\sigma]$-family contributions.
Second, among the $\sZ_2$-even operators, the leading ones 
are $T\bar{T}$ and $T^2 + \bar{T}^2$, both with $y=-2$. 
However, $T^2 + \bar{T}^2$ is a spin-four operator and thus it may contribute 
to rotationally invariant quantities only to second order, i.e. it 
gives corrections of order $\tau^4$. Therefore, the leading correction
(of order $\tau^2$) can only be due to $T\bar{T}$. Accordingly 
we write:
\begin{eqnarray}
\tilde{\phi} = - A \left(1 + 
          \phi_1 \mu^2_t \mu_{T\bar{T}} + O(\tau^4) \right), \\
\rho_h = B \left(1 + 
          \rho_{h1} \mu^2_t \mu_{T\bar{T}} + O(\tau^4) \right), \\
\psi_{\pm,hh} = C^\pm \left(1 + 
          \psi_{\pm,hh1} \mu^2_t \mu_{T\bar{T}} + O(\tau^4) \right). 
\end{eqnarray}
Then, since $\mu_{T\bar{T}}(\tau)$ is an even function of $\tau$, we have
for the functions $G_{\pm}(z)$ defined in \reff{hatF-square}
\be
G_\pm = 1 + (\psi_{\pm,hh1} - 2 \rho_{h1} + \phi_1) z^2 \mu_{T\bar{T}}(0) + 
            O(z^4).
\ee
By comparing with \reff{G-square}, we see that one of the following two
conditions must be satisfied: either 
$(\psi_{\pm,hh1} - 2 \rho_{h1} + \phi_1) = 0$ or $\mu_{T\bar{T}}(0) = 0$.
Thus, unless a miraculous cancellation occurs, the absence 
of the $z^2$ term implies our conjecture (d0).

Equation \reff{G-square} implies also that at least one operator contributes to 
order $\tau^4$ and a different one at order $\tau^6$. Note that it is 
not possible that the contribution of order $\tau^6$ is due to the 
nonlinear scaling field(s) already contributing to order $\tau^4$. 
Indeed, if this were the case, the contribution $O(z^6)$ in \reff{G-square}
would be independent of the phase as the term $O(z^4)$ is.\footnote{
Note that this independence does not follow from the RG 
expressions, since the functions $\psi_{+}$ and $\psi_{-}$ are expected to be different.}
This result is perfectly compatible with the CFT results of Sec. \ref{sec3} that predict:
\begin{enumerate}
\item At order $\tau^4$, the spin-four operator $T^2 +\bar{T}^2$ 
  appears;
\item At order $\tau^6$, three operators may appear: 
  the spin-zero operators $Q_4^{I} \bar{Q_4^{I}}$ and 
  $Q_3^{\sigma} \bar{Q_3^{\sigma}}$, and the spin-four 
  operator $Q_4^{\epsilon} + \bar{Q_4^{\epsilon}}$. 
\end{enumerate}
Note that $T^2 +\bar{T}^2$ and $Q_4^{\epsilon} + \bar{Q_4^{\epsilon}}$ have
$y=-2$ and $y=-3$ respectively; however, 
since they have spin four, they may contribute only at second order, 
and therefore at $O(\tau^4)$ and $O(\tau^6)$ respectively. 
Finally, note that \reff{G-square} is also in perfect agreement 
with the stronger conjecture (d2), that only non-rotationally
invariant operators are present. In this case, we have an operator 
that starts contributing at order $\tau^4$ and a second one appearing 
at order $\tau^6$. 

At higher orders, the situation becomes more involved. 
Beside the contributions of the expansion of the scaling fields appearing 
at lower orders, at order $\tau^8$ one must consider the fourth
power of the nonlinear scaling field associated to $T^2 +\bar{T}^2$. 
There is also a spin-zero operator 
$Q_4^{\epsilon}\bar{Q_4^{\epsilon}}$ with $y=-7$. However, because of 
the conjecture (b), we expect this operator 
to contribute only to second order and therefore starting at $O(\tau^{14})$.

It is interesting to note that, if the conjecture (d0) is true, Eqs. 
\reff{def-function-a} and \reff{def-function-b} provide the 
first terms of the expansion of $\mu_t(\tau)$ and $v_h(\tau)$ in 
powers of $\tau$. Explicitly
\begin{eqnarray}
\mu_t(\tau) &=& \tau \left( 1 - {3\over16} \tau^2 + O(\tau^4) \right), \\
v_h(\tau) &=& k(\tau)^{1/8} \left(1 + {11\over 128}  \tau^2 + O(\tau^4)\right).
\end{eqnarray}
Such expansions already appear in \cite{ssv1}, but assume a very simple 
form in the variable $\tau$. 

Finally, let us see which informations we can obtain from $B_f(\tau)$.
As we already noted our expressions are not compatible with 
\reff{Bf-def} because of the presence of higher powers of $\log \tau$.
We assume here
that our parametrization of the free energy gives the correct 
expression of $B_f(\tau)$ up to terms of order $\tau^{4}$, since at 
this order a $\log^2\tau$ appears. 
Under this assumption, we can compute the first terms in the 
expansion of $\lambda_t(\tau)$.  We compare the 
terms proportional to $\log |\tau|$, writing
\be
2 \mu_t(\tau) \lambda_t(\tau) \tilde{\phi}(\{0\}) = 
     \sum_{q=1}^3 b^{(1,q)} \tau^q + O(\tau^4).
\ee
Using $\tilde{\phi}(0)=-1/(2\pi)$, this gives for $\lambda_t(\tau)$
\be
\lambda_t(\tau) =\, k(\tau)^{1/4} \sum_{k=0}^\infty \lambda_{tk} \tau^k,
\label{lambdat-expansion}
\ee
where
\begin{eqnarray}
\lambda_{t0} &=& -0.10163764897527987657904520338506263625548489685 \; ,
\nonumber \\
\lambda_{t1} &=& 0 \, 
\nonumber \\
\lambda_{t2} &=& -0.000912698513043685863484370258366986546254622\; .
\end{eqnarray} 
It remains unclear why, by factoring out the term $k(\tau)^{1/4}$,
the linear term in $\lambda_t(\tau)$ vanishes. Note that the 
value of $\lambda_{t2}$ is correct only if the conjecture 
(d0) holds.

\subsection{The triangular lattice} \label{sec4.3}

It is very interesting to extend the results of \cite{g2000} to the 
triangular lattice. Indeed, in this case it is possible to make a much 
stronger test of the conjectures we have made. 

First, it is easy to see that the exact results 
\cite{Stephenson-64} for the free energy and 
the magnetization are fully compatible with the conjectures 
we have made. Then, 
let us derive the behavior of the susceptibility.
Equation \reff{RGprediction-chi} is lattice independent and 
it implies (apart from the logarithmic structure) \reff{Orrick-chi}. 
Therefore, the expansion on the triangular lattice should also have 
the form \reff{Orrick-chi}. Also, according to 
conjectures (a) and (b), we expect $\widehat{F}(\tau)$ to be even
in $\tau$, where now $\tau$ is defined in \reff{tau-tria}: 
some evidence will be provided in Sec. \ref{sec5.2}. 
Therefore, \reff{hatF-square} should hold with $G_\pm(z)$ even in $z$. 

Finally, we wish to predict which powers of $z$ should be absent in the 
expansion of $G_\pm(z)$. This depends on the operators that can appear.
CFT predicts the following:
\begin{enumerate}
\item At order $\tau^2$ we should consider $T\bar{T}$;
\item At order $\tau^6$ we should consider the spin-zero operators 
$Q_4^{I} \bar{Q_4^{I}}$ and $Q_3^{\sigma} \bar{Q_3^{\sigma}}$;
\item At order $\tau^8$ we should consider the spin-six operator
$Q^I_6 + \bar{Q}^I_6$; 
\item At order $\tau^{10}$ we should consider the spin-zero operators
 $Q_6^{I} \bar{Q_6^{I}}$, $Q_5^{\sigma} \bar{Q_5^{\sigma}}$,
 and the spin-six operators $Q_6^{\epsilon} + \bar{Q_6^{\epsilon}}$,
 $Q_6^\sigma + \bar{Q}_6^\sigma$.
\end{enumerate}
As we already mentioned, spin-six operators contribute to second order
in rotationally invariant quantities. Moreover, we have not indicated 
powers of lower-order operators and the 
$[\epsilon]$-family operator $Q^\epsilon_4\bar{Q}^\epsilon_4$ 
that, according to conjecture (b), should contribute corrections of 
order $\tau^{14}$. 

{} From this classification, we have the following 
possibilities: 
\begin{enumerate}
\item If $T\bar{T}$ is present, the term of order $z^2$ should be 
present barring miraculous cancellations.
\item If the conjecture (d0) is true, as on the square lattice, while 
the conjecture (d1) is false so that $\mu_{T\bar{T}}(\tau) \sim \tau^2$, 
then the term of order $z^2$ should be absent and the term of 
order $z^4$ should be nonvanishing.
\item If the conjecture (d1) is valid, 
both terms of order $z^2$ and $z^4$ should be absent;
\item If the stronger conjecture (d2) is true, i.e. if only 
non-rotationally invariant operators are present, the term 
of order $z^6$ is also absent. 
More precisely, this cancellation would imply 
$\mu(0)= 0$ for $Q_4^I \bar{Q}_4^I$, $v(0) = 0$ for 
$Q_3^\sigma \bar{Q}_3^\sigma$, and $\mu_{T\bar{T}}(\tau)\sim o(\tau^4)$.
We expect the term of order $z^8$ to be nonvanishing
since at this order the spin-six operator 
$Q_6^I + \bar{Q}_6^I$ should contribute.
\end{enumerate}

The triangular lattice is therefore a better testing ground for our 
conjectures. Indeed, the conjecture (d1) requires two coefficients to vanish, 
a very nontrivial fact. Moreover, we are able to distinguish between 
the conjectures (d1) and (d2).

\section{The large-distance behavior of the two-point function} \label{sec5}

In this Section we will study the large-distance behavior of the two-point
function on the square lattice, reviewing in part the 
results of \cite{CCCPV-00}, and on the triangular lattice.
The square-lattice analysis will confirm the validity of the 
conjecture (d0), i.e. $\mu_{T\bar{T}} (0) = 0$.
Much more interesting is the analysis on the triangular lattice 
which will show that $\mu_{T\bar{T}} (\tau) = o(\tau^4)$, thus providing 
strong support to the conjecture (d1). We will also find that the 
subleading corrections due to the zero-spin operator with 
$y=-6$ are absent, in agreement with the conjecture presented in the 
Introduction (conjecture (d2) of Sec. \ref{sec4.1}).

\subsection{The square lattice} \label{sec5.1}

Let us now consider the large-distance behavior of the two-point function 
for $h=0$, $\tau > 0$. For large $|x|$ it has the form 
\cite{CW-67} 
\be
G(x,y;\tau) = Z(\tau) \int_{-\pi}^\pi {dk_1\over 2\pi} {dk_2\over 2\pi}
        {e^{ik_1 x+ ik_2 y} \over \Delta_s(k) + M_s(\tau)^2} ,
\ee
where 
\begin{eqnarray}
  \Delta_s(k) &=& 4 \sin^2 {k_1\over2} +\ 4 \sin^2 {k_2\over2}, \\
  Z(\tau) &=& \sqrt{8}\, \tau^{1/4}\, k(\tau)^{1/4}\, (1 + \tau^2)^{1/8} = 
       2 (k(\tau)^2 - 1)^{1/4},  
\label{Z-square}\\
  M_s(\tau)^2 &=& 4 \left(\sqrt{1 + \tau^2} - 1\right).
\end{eqnarray}
{} From these expressions, we can compute the angle-dependent correlation 
length $\xi(\theta)$ defined from the large-distance behavior of the 
two-point function along a direction forming an angle $\theta$ with 
the side of the lattice. We obtain
\be
\xi(\theta) = {1\over\sqrt{2}a(\tau)} 
   \left[1 + {a(\tau)^2\over 48} \cos4\theta + 
      {a(\tau)^4} \left({1\over 3072} - {1\over 320} \cos 4\theta - 
        {5\over 9216} \cos 8 \theta\right) + O(a(\tau)^6)\right],
\label{xi-square}
\ee
where $a(\tau)$ is defined by Eqs. \reff{def-function-a}, \reff{a-function-square}.
As already observed in \cite{CPRV-98}, this expansion 
shows the presence of a correction of order $\tau^2$ due to the leading 
irrelevant operator breaking rotational invariance. 
However, the interesting additional feature is that this term is the 
only one, i.e. there is no correction due to the rotationally 
invariant subleading operators \cite{CCCPV-00}. 
This result is naturally interpreted: The correction we find is due to 
the spin-four operator $T^2 + \bar{T}^2$ and there is no contribution 
due the scalar operator $T\bar{T}$. At order $\tau^4$ there is 
scalar term, but this does not require the presence of a scalar operator: 
The angle-independent contribution  can be interpreted as due to the 
square of the spin-four operator $T^2 + \bar{T}^2$. 
Therefore, the result \reff{xi-square} supports the conjecture (d0)
and is compatible with the stronger ones (d1) and (d2).

In \cite{CCCPV-00} we also analyzed the on-shell renormalization constant
$Z(\tau)$ and found no terms of order $\tau^2$. We thought this to be a 
good indication of the absence of both $T\bar{T}$ and $T^2 + \bar{T}^2$.
We now believe that this conclusion was a little bit too hasty. 
First, \reff{Z-square} implies
\be
Z(\tau) = \sqrt{8} a(\tau)^{1/4} b(\tau)^2,
\label{Zsquare}
\ee
with no corrections to all orders. Of course, we cannot take this 
as an indication that all operators are absent. 
Moreover, there is also a conceptual problem: $Z(\tau)$ is defined from the 
behavior of the two-point function at $p = - i M(\tau)$ and thus 
we should consider the momentum-dependent nonlinear scaling fields as 
we did in \cite{CCCPV-00} for the second-moment correlation length.
As we shall see in the next Section, no particular simplification occurs in the 
triangular case, and we find corrections of order $\tau^2$ to the 
expression \reff{Zsquare}. Thus, the observed cancellation is accidental
and does not have any connection with the operator structure of the model.

Finally, we present an argument to make plausible the fact that the 
functions $\widehat{F}_\pm(\tau)$ are even in $\tau$. 
If the short-distance part $B_f(\tau)$ were absent, such a property 
would follow from the symmetry
\be
  (-\tau)^{-1/4}\,  k(-\tau)^{-1/4} \chi_\pm(-\tau) = 
    \tau^{-1/4} \, k(\tau)^{-1/4} \chi_\pm(\tau).
\ee
The interesting observation is that this symmetry property is satisfied by
the large-distance expression of $G(x,y;\tau)$. Indeed, using the 
expressions reported above we immediately verify that
\be
  (-\tau)^{-1/4}\, k(-\tau)^{-1/4} G(x,y;-\tau) = 
  \tau^{-1/4} \, k(\tau)^{-1/4}  G(x,y;\tau).
\label{relation-G-dualita}
\ee

\subsection{The triangular lattice} \label{sec5.2}

We now repeat the same analysis on the triangular lattice. The large-distance 
behavior of the two-point function along a side of the lattice was 
computed in \cite{Stephenson-70}. Such expression was generalized in
\cite{CPRV-96} where it was conjectured that the large-distance behavior 
was given by the propagator of a Gaussian field on a triangular 
lattice, in analogy with the square-lattice expression. 
Therefore,
\be
G(x,y;\tau) = {\sqrt{3}\over 8\pi^2} \, Z(\tau) \int^\pi_{-\pi} {dk_1}
           \int^{2 \pi/\sqrt{3}}_{-2 \pi/\sqrt{3}} dk_2\,
            {e^{i k_1 x + i k_2 y}\over \Delta_t(k) + M_t(\tau)^2},
\label{generalG-tr}
\ee                                                                             
where 
\begin{eqnarray}
\Delta_t(k) &=& 4 - {4\over3} \cos k_1 -
                  {8\over3} \cos {k_1\over2} \cos {\sqrt{3} k_2\over2} ,\\
M_t(\tau)^2 &=& {8\over3} \left(\cosh \textstyle{1\over2}\mu_l - 1\right)
                    \left(\cosh \textstyle{1\over2}\mu_l + 2\right)  , \\ 
Z(\tau) &=& {8\over 3} A(\tau)^{-1/4} (k(\tau)^2 - 1)^{1/4} 
            \left(A(\tau) + \sqrt{A(\tau)} + 1\right)^{1/2}, \\
\mu_l(\tau) &=& \log A(\tau), 
\end{eqnarray}
and 
\be
A(\tau) \equiv 
\left({\sqrt{1 - v + v^2} - \sqrt{v} \over \sqrt{v} (1 - v)}\right)^2.
\ee 
The conjectured form \reff{generalG-tr} was checked in the 
high-temperature limit \cite{CPRV-96}, by computing the 
expansion of $G(x,y;\tau)$ in powers of $\beta$ to order $\beta^{15}$.

Note that, under $\tau \to -\tau$, we have 
\be
A(- \tau) = {1\over A(\tau)},
\ee
and 
\begin{eqnarray}
M_t(-\tau)^2 &=& M_t(\tau)^2,  
\label{duality-Mt} \\
Z(-\tau) (-\tau)^{-1/4} k(-\tau)^{-1/4} &=& 
     Z(\tau) \tau^{-1/4} k(\tau)^{-1/4}.
\label{duality-Zt}
\end{eqnarray}
{}From the large-distance behavior of the two-point function we can obtain 
the angle-dependent correlation length $\xi(\theta)$ taken along 
a direction forming an angle $\theta$ with a side of the triangles. 
We have, in terms of the function $a(\tau)$ defined in 
\reff{def-function-a}, \reff{a-function-tria}, 
\begin{eqnarray}
\xi(\theta) &=& {\sqrt{3} \over 2 a(\tau)} 
    \left[1  + {a(\tau)^4 \cos 6\theta\over 6480} - 
               {a(\tau)^6 \cos 6\theta\over 54432} \right.
\nonumber \\
  && \hphantom{\sqrt{3} \over 2 a(\tau)} 
  \left. + {a(\tau)^8\over 55987200} +
   {a(\tau)^8 \cos 6\theta\over 559872} - 
   {a(\tau)^8 \cos 12\theta\over 18662400} \right].
\end{eqnarray}
This result provides a very strong check of the conjecture (d2) presented in 
the introduction. Indeed, the first correction term appears only at order 
$a(\tau)^4$ and is proportional to $\cos 6\theta$. It is thus unambiguously 
related to the spin-six operator $T^3 + \bar{T}^3$. At order $a(\tau)^6$ 
there is also a correction term, but it is again proportional to 
$\cos 6\theta$ and thus it should be associated to a spin-six 
operator. Since no new operator appears at this order, it must be identified
with an analytic correction due to the operator 
$T^3 + \bar{T}^3$. At order $a(\tau)^8$ a constant term and 
a $\cos 12\theta$ appear,
but they may be due to the square of the operator $T^3 + \bar{T}^3$.

In conclusion, this calculation provides very strong evidence for the 
absence of $T\bar{T}$, conjecture (d1)---more precisely it proves that
$\mu_{T\bar{T}} = o(\tau^4)$---and also for the
conjecture (d2). Indeed, if (d1), but not (d2), were true, the 
spin-zero operator $Q_4^I + \bar{Q}_4^I$
would contribute to order $\tau^6$, giving rise to an angle-independent term 
proportional to $a(\tau)^6$. The absence of such term supports  the validity 
of (d2).

Interestingly enough, this calculation allows the 
computation of the first analytic term in the scaling field $\mu_1(\tau)$ 
that is associated with $T^3 + \bar{T}^3$. Indeed, if the conjecture (d2) is 
correct, the function $a(\tau)$ 
given in \reff{a-function-tria} coincides with the temperature scaling 
field at $h=0$ up to terms of order $\tau^9$, i.e. 
$\mu_t(\tau) = a(\tau) + O(\tau^9)$. Then, we write 
\be
\xi(\theta) = {\sqrt{3}\over 2} {1\over \mu_t(\tau)}
   \left(1 + \alpha \mu_t(\tau)^4 \mu_1(\tau) \cos 6\theta + O(\tau^8)\right),
\ee
and fix $\alpha$ by requiring $\mu_1(0) = 1$. Then 
\be 
\mu_1(\tau) = 1 - {5\over 42} \tau^2 + O(\tau^4).
\ee
Considering now the function $Z(\tau)$, no particular simplification
occurs and a correction term of order $a(\tau)^2$ 
appears. Explicitly
\be
Z(\tau) = {16\over 3\cdot 6^{1/4}} a(\tau)^{1/4} b(\tau)^2 
\left(1 + {a(\tau)^2\over 18} + \cdots\right).
\ee
As we already discussed in Sec. \ref{sec5.1},
the presence of the quadratic term is probably related to the 
presence of a momentum-dependent contribution to the 
nonlinear scaling fields.

Finally, we note that \reff{relation-G-dualita} is also satisfied 
on the triangular lattice, as it may be easily shown
by using \reff{duality-Mt} and \reff{duality-Zt}. 
Again, this gives a plausibility argument for the 
fact that the function $\widehat{F}(\tau)$ appearing 
in \reff{Orrick-chi} is even on the 
triangular lattice too.

\section{Finite-size scaling at the critical point} \label{sec6}

Recently, there has been much effort in understanding the behavior of the 
Ising model in a finite box or strip of size $L$ at the critical point 
$h=\tau=0$, computing the finite-size free energy $f_L$, 
energy $E_L$, specific heat $C_L$, and inverse mass gap 
$\xi_L$. The results obtained are the following:
\begin{itemize}
\item In \cite{deQueiroz} and \cite{ih-00-strip}, $f_L$ and $\xi_L$
were computed on a strip of width $L$ for several different lattices: 
It was found that these two quantities have an expansion of the form 
\begin{eqnarray}
L^2 (f_L - f_\infty) &=& \sum_{n=0}^\infty {f_n\over L^{2n}} \\
{\xi_L\over L} &=& \sum_{n=0}^\infty {s_n\over L^{2n}}.
\end{eqnarray}
Note that in the expansion only even powers of $L$ appear. Moreover, 
on a triangular lattice $f_1=f_3=0$ and $s_1=s_3=0$.
\item Salas \cite{Salas-01} and Izmailian and Hu \cite{ih-00-square} 
computed $f_L$, $E_L$, $C_L$ for a square lattice $L\times M$
for fixed aspect ratio $\rho = M/L$, 
extending the results of \cite{FeFi}. They found:
\begin{eqnarray}
L^2 (f_L - f_\infty) &=& \sum_{n=0}^\infty {f_n(\rho)\over L^{2n}}, \\
E_L &=& - \sqrt{2} + \sum_{n=0}^\infty {e_n(\rho)\over L^{2n+1}}, \\
C_L &=& {8\over \pi} \log L + \sqrt{2} E_L + 
    \sum_{n=0}^\infty {h_n(\rho)\over L^{2n}}. 
\end{eqnarray}
The specific heat has also been computed for a square lattice 
with Brascamp-Kunz boundary conditions in \cite{JK-01}.
However, in this case translation invariance is lost 
in one direction and thus we cannot apply straightforwardly the 
results presented here.
\end{itemize}
In this Section, we want to explain the general features of these
results. 

In finite volume the general scaling expression \reff{F-RG} can be 
generalized by writing (see, e.g., \cite{PriRu,GJ-87,PHA-91,Privman-90})
\be
F(\tau,h;L) = f_b(\tau,h) + {1\over L^2} W(\{u_j L^{y_j}\}) + 
       {1\over L^2} \log L\, \widetilde{W}(\{u_j L^{y_j}\}),
\label{F-RG-finiteL}
\ee
where we assume that the bulk contribution is independent of $L$, or,
more plausibly, that it depends on $L$ only through exponentially small 
corrections \cite{Privman-90,PHA-91}, and the functions $W$ 
and $\widetilde{W}$ depend on all 
scaling fields. Equation \reff{F-RG-finiteL} cannot be correct in general. 
Indeed, the results of \cite{g2000} indicate the presence of powers of
$\log|\tau|$ in the susceptibility, which imply the 
presence of powers of $\log L$ in \reff{F-RG-finiteL}. 
These corrections should be relevant only if we consider 
derivatives of the free energy with respect to $h$, while here 
we set $h=0$ from the beginning. In this particular case, 
\reff{F-RG-finiteL} should be correct. 

If $h=0$, the 
$[\sigma]$-family scaling fields do not contribute, 
so that \reff{F-RG-finiteL} becomes
\be
F(\tau,0;L) = f_b(\tau,0) + {1\over L^2} W(\{\mu_j(\tau) L^{y_j}\}) +
       {1\over L^2} \log L\;  \widetilde{W}(\{\mu_j(\tau) L^{y_j}\}),
\label{F-RG-finiteL-h0}
\ee                                                                             
where the scaling functions depend only on the ${\sZ}_2$-even scaling fields.
By using \reff{scaling-fields-duality} and the fact that 
the RG eigenvalues $y_j$ are even for the identity family and 
odd for the energy family we obtain
\be
  W(\{\mu_j(-\tau) (-L)^{y_j}\}) =  W(\{\mu_j(\tau) L^{y_j}\})
\ee
and an analogous formula for $\widetilde{W}$. 
Therefore, apart from the bulk contribution,
even derivatives of $F$ with respect to $\tau$ contain only even powers of 
$L$, while odd derivatives contain only odd powers of $L$. This explains 
the particular structure of the results obtained by 
\cite{ih-00-strip,ih-00-square,Salas-01} since
\begin{eqnarray}
E_L &=& 2 \sqrt{2} \left. {\partial F\over \partial\tau}\right|_{\tau = 0},
\\
C_L &=& \sqrt{2} E_L + 8 
  \left. {\partial^2 F\over \partial\tau^2}\right|_{\tau = 0}.
\label{RG-prediction-CL}
\end{eqnarray}
In particular, \reff{RG-prediction-CL} explains why
the odd terms in the expansion of $C_L$ are related 
to those of the  energy. 

For what concerns the logarithms, only $C_L$ shows a logarithmic 
dependence, and only at leading order in $L$. This may be explained
if
\be
 \widetilde{W}(\{\mu_j(\tau) L^{y_j}\}) = 
   \widehat{W}(\mu_t(\tau) L).
\label{eq:What}
\ee
By using the results for the specific heat at criticality and in the 
infinite-volume limit we can compute the asymptotic behavior of 
$\widehat{W}(x)$ for $x\to 0$ and $x\to \infty$. 
For $x\to 0$, the results for $C_L$ imply
\be
\widehat{W}(x) \approx {1\over 2\pi} x^2 + O(x^4),
\ee
while in order to obtain the correct infinite-volume limit, we should have
\be
\widehat{W}(x) \approx {1\over 2\pi} x^2 \left(1 + O(x^{-2})\right).
\ee
These two results make natural the conjecture that 
\be 
\widehat{W}(x) = {1\over 2\pi} x^2 
\label{eq2:What}
\ee
for all $x$. There are several consequences of these results: 
\begin{itemize}
\item 
Relation \reff{eq:What} and conjecture (c1) imply conjecture (c2), i.e. 
that the function 
$\tilde{f}$ in \reff{F-RG} is a simple constant, as originally suggested
by Aharony and Fisher \cite{af}. If this is the case, the function 
$\mu_t(\tau)$ coincides with the function $a(\tau)$.
\item 
If \reff{eq2:What} is correct, we predict that in the expansion of 
$\partial^{2n} F/\partial \tau^{2n}$ at the critical point 
there is only one logarithmic term, with a coefficient that can be computed 
from the expansion of $a(\tau)$.
\end{itemize}
Let us now use \reff{F-RG-finiteL-h0} to determine the 
corrections to the leading behavior. 
We obtain
\begin{eqnarray}
L^2 f_L &=& L^2 f_b(0,0) + W(\{x_j\}), 
\label{f-RG-L}\\
{\partial F\over\partial\tau}(0) &=& 
   \left.{\partial f_b\over \partial \tau}\right|_{\tau=h=0} + 
        {1\over L^2} \sum_{i\in [\epsilon]} L^{y_i} W_{i}(\{x_j\}), 
   \\
{\partial^2 F\over\partial\tau^2}(0) &=& 
     \left. {\partial^2 f_b\over \partial \tau^2}\right|_{\tau=h=0}  + 
          {1\over L^2} \sum_{ik\in [\epsilon]}  
                         L^{y_i+y_k} W_{ik} (\{x_j\}), 
\nonumber \\ 
         && + {1\over L^2} \sum_{i\in [I]} \mu_{2,i} L^{y_i}
               W_{i} (\{x_j\}) + 2 A \log L,
\end{eqnarray}
where we write $\mu_j(\tau) = \mu_j(0) + \tau \mu_{1,j} + 
{1\over2} \tau^2 \mu_{2,j}$, the functions $W_{i}$, and 
$W_{ik}$ depend only on the identity-family scaling fields through
$x_j \equiv \mu_j(0) L^{y_j} = L^{y_j}$, and the constant $A$ is defined by
\reff{def-function-a}. We have also used the normalization conditions
$\mu_{1,i} = 1$ for the energy-family fields 
and $\mu_j(0) = 1$ for the identity-family fields.

Let us now discuss which corrections should be expected. The important point 
is that here, at variance with the infinite-volume case, 
nonzero spin operators can contribute to first order. Indeed, the 
box breaks the rotational invariance down to the lattice invariance 
and therefore the mean value of a lattice operator 
that is not rotationally invariant but has the symmetries of the lattice is 
nonzero. This implies that no missing term is expected on the 
square lattice, in agreement with the exact results. 
Indeed, the lowest operator is the spin-four operator $T^2+\bar{T}^2$ that 
has $y=-2$ and belongs to the identity family, and is therefore able, alone,
to give rise to all observed corrections.

On the triangular lattice instead simplifications are expected. 
Consider first, the free energy $f_L$. The absence of the term 
proportional to $L^{-2}$, i.e. $f_1 = 0$, implies 
$\mu_{T\bar{T}}(0) = 0$, confirming once again the 
conjecture (d0). 
The next-to-leading operator belonging  to the identity family is 
the spin-six $T^3+\bar{T}^3$
that has $y=-4$. Therefore, in \reff{f-RG-L} the $T^3+\bar{T}^3$ 
gives rise to corrections of order $L^{-4n}$. 
The absence of the 
$1/L^6$ term requires an additional cancellation, 
i.e. $\mu(0)$ for the operator $Q_4^I \bar{Q}_4^I$ that has $y=-6$ 
and zero spin, thereby 
supporting our conjecture (d2).
At order $1/L^8$ there appears a new operator 
$Q_2^{I}\bar Q_8^{I}+\bar Q_2^{I} Q_8^{I}$
that gives, together with $T^3 + \bar{T}^3$,
corrections of order $L^{-8n-4m}$ and thus 
indistinguishable from those of $T^3+\bar{T}^3$. 
At order $1/L^{10}$, at least the spin-12 operator $T^6 + \bar{T}^6$ appears 
and therefore we expect all corrections of the form $L^{-10n-4m}$ to be 
nonvanishing.

An analogous cancellation is expected for $E_L$. 
For $E_L$ the leading correction terms are 
\be
{1\over L} \mu_{1,t} W_t(\{x_j\}) + 
{1\over L^7} \mu_{1,1} W_1(\{x_j\}) + \ldots
\ee
where $\mu_{1}(\tau)$ is the scaling field of the spin-six
operator $Q^\epsilon_6 + \bar Q^\epsilon_6$ that has $y=-5$. 
Reasoning as before, on the basis of conjecture (d0) alone, 
we expect no correction of order $1/L^3$ 
but the presence of all other terms. Analogously in $C_L$ the 
$L^{-2}$ correction should be absent.

The results for the correlation length show the same pattern of the 
free energy. The fact that $s_1 = s_3 = 0$ on the triangular lattice 
provides additional evidence for the absence of spin-zero operators 
in the theory.

It is interesting to notice that a similar finite-size scaling analysis was
performed more than 10 years ago for the one-dimensional Ising quantum chain
which belongs to the same universality class of the two-dimensional 
Ising model (for a
discussion of their connection, see~\cite{bg87}). In particular,
in~\cite{h87} the finite-size behavior of the free energy and of the mass
spectrum of the model was obtained and then compared  in ~\cite{r87a,r87b} 
with the predictions of perturbed CFT (see~\cite{H_book} for an 
updated review of the subject).

Remarkably enough, also in this case the contribution of the energy-momentum
tensor exactly disappears and the first non-zero correction is given again by
the spin-four operator $T^2 +\bar T^2$~\cite{r87a}.

\section{Finite-size scaling of the susceptibility at $t=0$} \label{sec7}

In the previous section we have discussed several thermal quantities at the 
critical point and verified that the observed behavior is consistent with 
the RG and CFT predictions and the conjectures we have made. 
Here, we want to discuss the finite-size behavior of the susceptibility 
on the square lattice, and we will check that the correction
coefficients depend on the shape of the domain as predicted by the 
spin nature of the operators. 

For this purpose we study two different finite square lattices in order 
to verify the dependence of the corrections on the domain:
\begin{eqnarray}
D^{(A)}_M &=& \left\{ (n_0,n_1)\in \sZ^2, \; 0\le n_1,n_2\le M-1 \right\},
\\
D^{(B)}_M &=& \left\{ (n_0,n_1)\in \sZ^2, \; 0\le n_1+n_2\le 2M-1, 
      0\le n_1-n_2\le 2M-1
       \right\}.
\end{eqnarray}
In both cases the domain is a square: the first one
has boundaries that are parallel to the lattices axes and 
size $L=M$, while the 
second one is rotated by 45$^{o}$ and has size $L=M\sqrt{2}$. 
We use periodic boundary conditions. For domain $(A)$ 
such conditions are obvious, for domain $(B)$ we identify 
$(n_1,n_2)$ with $(n_1+M,n_2+M)$ and $(n_1+M,n_2-M)$.

\subsection{Renormalization-group analysis}
\label{sec7.1}

The finite-size scaling behavior of the susceptibility can be derived 
easily, starting from \reff{F-RG-finiteL}. As we already said, such an 
expansion misses some important corrections proportional
to higher powers of $\log L$. However, they should only 
be of interest if we analyzed the asymptotic behavior
of $\chi$ for $\tau\to0$. Here, we consider 
$\chi$ at the critical point and thus such corrections should 
vanish.

A simple computation gives at the critical point 
\begin{eqnarray}
\chi_L(0,0) &=&
  \left. {\partial^2 f_b\over \partial h^2}\right|_{\tau=h=0} + 
  {1\over L^2} \sum_{k\in\, [I],[\epsilon]} \lambda_k(0) L^{y_k} 
        W_k(\{x_j\}) 
  \nonumber \\
  && + {1\over L^2} \sum_{ik\in\, [\sigma]} 
  L^{y_i + y_k} W_{ik}(\{x_j\}),
\end{eqnarray}
where the functions depend only on the identity-family scaling fields,
$x_j \equiv \mu_j(0) L^{y_j} = L^{y_j}$,
and we have used the normalization conditions 
$v_i(0) = 1$, $\mu_j(0) = 1$ for spin- and identity-family 
scaling fields.

Since $y_j = 2n-\frac18$ for the $[\sigma$]-family operators and 
$y_j=2n$ for the identity-family operators, where $n$ is an integer,
we have
\be
{1\over L^2} \sum_{ik\in\, [\sigma]} 
  L^{y_i + y_k} W_{ik}(\{x_j\}) = 
  L^{7/4}\, \sum_{k=0}^\infty {c_k\over L^{2k}},
\ee
i.e. the corrections contain only even powers of $L$. On the square lattice
we do not anticipate any cancellation, i.e. we expect $c_k\not=0$ for all 
$k$. Indeed, the leading correction is due to the operator
$T^2 + \bar{T}^2$, which has $y=-2$, and thus gives rise to corrections 
involving all powers of $L^{-2}$. 
On the triangular lattice instead we expect $c_1 = 0$, because of 
the conjecture (d0). All other terms are expected to be 
nonvanishing. Indeed, the presence of the spin-six operator
$T^3 + \bar{T}^3$ generates terms $L^{-4n}$, while 
the presence of the spin-six operator $Q^\sigma_6 + \bar{Q}^\sigma_6$
together with the previous one generates terms $L^{-6-4n}$. 

Let us now consider the term that contains a sum over all 
identity- and energy-family operators. We 
expect in this case all powers of $L^{-1}$, i.e.
\be
{1\over L^2} \sum_{k\in\, [I],[\epsilon]} \lambda_k(0) L^{y_k} 
    W_k (\{x_j\}) = {1\over L}
       \sum_{k=0}^\infty {d_k\over L^{k}}.
\label{sviluppochiL-secondo}
\ee
On the square lattice we should have $d_1 = 0$. Indeed, the leading 
energy-family scaling field is associated with the temperature and gives a 
contribution of the form 
\be
{1\over L^2} \lambda_t(0) L W_t (\{x_j\}) \sim {1\over L} 
   \left(a + {b\over L^2} + 
        {c\over L^4} + \cdots\right),
\ee
and thus generates all even terms in \reff{sviluppochiL-secondo}.
The odd terms in \reff{sviluppochiL-secondo} are generated by the 
identity-family operators, the leading one being $T^2 + \bar{T}^2$. 
It gives 
\be
{1\over L^2} \lambda_1(0) L^{-2} W_1 (\{x_j\}) \sim {1\over L}
   \left( {a\over L^3} + {b\over L^5} + {c\over L^7} + \cdots 
   \right),
\ee
and thus generates all odd terms except the first one. Hence $d_1=0$. 
Note that is cancellation follows from CFT alone and does not require
any additional hypothesis.

On the triangular lattice the discussion is similar although a little 
more complicated. We predict $d_1 = d_2 = d_3 = d_7 = 0$. 
The condition $d_1=0$ does not require any conjecture, while $d_2=0$ 
implies the validity of the conjecture (d0). Much more interesting 
is to check whether $d_3=d_7=0$, since the vanishing of these coefficients
implies $\lambda_{T\bar{T}}(0)=0$ and 
$\lambda(0) = 0$ for the operator $Q_4^I \bar{Q}_4^I$. 
Thus, the analysis of $\chi$ on the 
triangular lattice would provide some additional evidence for or rule 
out the conjectures (d1) and (d2).

\bigskip 
\subsection{The transfer-matrix calculation} \label{sec7.2}

{} From the previous discussion, we can write on the square lattice 
\eq
 \chi_L(0,0) = L^{7/4}\left(c_0+\frac{c_1}{L^2}+\frac{c_2}{L^4}\right)
 + D_0 + L^{-1} \left(d_0 + {d_2\over L^2} + 
    {d_3\over L^3} \right) + O(L^{-17/4}, L^{-5}).
\label{eq5.2}
\en
The constant $D_0$ is lattice and geometry independent being generated 
by the bulk free energy, and it is given by $B_f(0)$. 
Explicitly:
\eq
 D_0 = B_f(0) \approx - 0.104133245093831026452160126860473433716236727314
\en
The other constants depend on the geometry of the system and in general 
are expected to be different for the two domains (A) and (B). However, 
this should depend on the type of operator that generates them. 
If a term is associated with a spin-zero operator, its value should be identical 
in both geometries, while if it is the first contribution of a spin-four
operator we expect a dependence of the form $\cos 4\theta$, where 
$\theta$ is the angle between the boundaries of the 
domain and the lattices axes. For our specific case, 
since $\theta=\frac\pi4$ we expect the 
coefficient to change sign. Therefore, we predict 
\be
c_0^{A} = c_0^B, \qquad c_1^A = - c_1^B, \qquad d_0^A = d_0^B.
\label{prediction-chiL}
\ee
Indeed, $c_0$ and $d_0$ are related to the magnetic and to the thermal scaling 
fields that have both spin zero. On the other hand, $c_1$ is related to the 
leading identity-family operator with $y=-2$. If the conjecture (d0) 
is correct, this term should be due only to the spin-four operator
$T^2 + \bar{T}^2$ and thus, according to the previous 
discussion, it should differ by a sign in the two geometries.

In the following we shall test the predictions \reff{prediction-chiL}. 
For this purpose it is interesting to note that the constants $d_0^A$ 
and $d_2^A$
can be predicted by using the results of 
\cite{FeFi,ih-00-square,Salas-01}. Indeed, 
\be
\lambda_t(0) W_t(\{x_j\}) = d_0 + {d_2\over L^2} + O(L^{-3}),
\ee
since the leading irrelevant operator contributing to 
\reff{sviluppochiL-secondo} has $y = -2$. 
Now, $\lambda_t(0)$ is given in \reff{lambdat-expansion}, while the 
leading contributions to the left-hand side can be derived from the 
energy at the critical point, since
\be
E_L = 2 \sqrt{2} {\partial F\over \partial \tau}(0) = 
   \left. 2 \sqrt{2} {\partial f_b\over \partial \tau}\right|_{\tau=h=0} + 
   {2 \sqrt{2}\over L} W_t(\{x_j\}) + O(L^{-5}).
\ee
For geometry (A), using the results of \cite{FeFi,ih-00-square,Salas-01}, 
we have
\be
W_t(\{x_j\}) = w_{t1} + {1\over L^2} w_{t2} + O(L^{-4}), 
\ee
where
\begin{eqnarray}
w_{t1} &=& - {1\over \sqrt{2}} 
  {\theta_2(0) \theta_3(0) \theta_4(0) \over 
   \theta_2(0) + \theta_3(0) + \theta_4(0)} 
  \approx - 0.220065581798270538286514481651 
\\
w_{t2} &=& {\pi^3 \over 96 \sqrt{2}}
  {\theta_2(0) \theta_3(0) \theta_4(0) 
  [\theta_2(0)^9 + \theta_3(0)^9 + \theta_4(0)^9]
  \over
   [\theta_2(0) + \theta_3(0) + \theta_4(0)]^2} 
\nonumber  \\ [2mm]
   &\approx& 0.073073526812330794515803384757
\end{eqnarray}
so that 
\begin{eqnarray}  
d_0^A &\approx& 0.022366948354353361434648349198, 
\label{C1A-theory} \\
d_2^A &\approx&  -0.007427021467537379563283082599.
\label{d2A-theory}
\end{eqnarray}
Note that this calculation relies only on the RG and on the 
CFT classification of the operators, but does not make use 
of any of the additional conjectures.

In order to check Eqs. \reff{eq5.2} and \reff{prediction-chiL},
we performed a transfer-matrix (TM) calculation of the susceptibility.
Notice that in general it is more difficult to perform a TM
calculation in the case in which both sizes of the lattice are finite than in
the case in which one of them is infinite, since one has to keep into 
account all the eigenvalues of the TM. 

\subsubsection{Numerical results} \label{sec7.2.1}

Let us see in detail the two cases that we
studied:
\begin{itemize}
\item
{\bf Geometry (A)}

In this case we computed $\chi$ on lattices of sizes up to $L=17$. 
In order to test our
methods we  evaluated the susceptibility in two ways, by direct differentiation
of the free energy and by using the fluctuation-dissipation theorem, i.e. by
summing the two-point function. The results are
reported in Table \ref{latA}. 
By comparing the two columns one can estimate the size of the
systematic errors.

\begin{table}
\begin{center}
\caption{\sl 
Numerical estimate of the magnetic susceptibility for
 geometry (A).
 In the second column we give the results obtained by 
 differentiation of the free energy
 and in the third 
 column those obtained by summing 
 the time-slice two-point correlation function.}
\vskip0.5cm
\label{latA}
\begin{tabular}{|r|l|l|}
\hline
\multicolumn{1}{|c|}{$L$} &  \multicolumn{1}{c|}{$\chi$}   &   
\multicolumn{1}{c|}{$\chi$}  \\
\hline
4 & \phantom{1}12.181742537099 & \phantom{1}12.18174253709876  \\
5 & \phantom{1}18.092431830874 & \phantom{1}18.09243183087397  \\
6 & \phantom{1}24.959397280867 & \phantom{1}24.95939728086672  \\
7 & \phantom{1}32.740662899119 & \phantom{1}32.74066289911872  \\
8 & \phantom{1}41.402340799629 & \phantom{1}41.40234079963127  \\
9 & \phantom{1}50.915891978613 & \phantom{1}50.91589197861391  \\
10& \phantom{1}61.256768274856 & \phantom{1}61.25676827485805  \\
11& \phantom{1}72.403538830976 & \phantom{1}72.40353883097585  \\
12& \phantom{1}84.337262930730 & \phantom{1}84.33726293072681  \\
13& \phantom{1}97.041023059667 & \phantom{1}97.04102305966430  \\
14& 110.49957085440 & 110.4995708543933 \\
15& 124.69905432425 & 124.6990543242478 \\
16& 139.62680432571 & 139.6268043257091 \\
17& 155.27116484686 & 155.2711648468523 \\
\hline
\end{tabular}
\end{center}
\end{table}

\item
{\bf Geometry (B)}

In order to study geometry (B) we used the following trick.
As a first step, we performed a decimation of the lattice, i.e. every 
second spin was integrated out. In 
this way the number of spins is reduced by half.
The price one has  to pay is  that the Hamiltonian becomes
more complicated  and contains, in addition to
 the nearest-neighbour interaction, a next-to-nearest neighbour and
 a four-point interaction. 
In the presence of an external field also a three-point
term arises. 

However, now the axes of the decimated lattice are parallel 
to the axes of the torus.
Also, the new Hamiltonian only couples neighboring time slices. 
Therefore, we can 
apply the same TM methods used in geometry (A).

Our numerical results are given in Table 
\ref{latB}.
We computed the magnetic susceptibility by 
differentiation of the free energy. 
The largest lattice has $M = 12$, which corresponds to $L=16.98$, and is thus 
completely equivalent to the largest lattice used in geometry (A).

\begin{table}
\caption{\sl 
Numerical result for the inverse of the magnetic susceptibility for 
geometry (B).}
\vskip0.5cm
\label{latB}
\begin{center}
\begin{tabular}{|r|l|}
\hline
\multicolumn{1}{|c|}{$M$}& \multicolumn{1}{|c|}{$1/\chi$}  \\
\hline
 2&0.149678741567431 \\
 3&0.073301790137056 \\
 4&0.044241139068172 \\
 5&0.029917172878427 \\
 6&0.021735601983740 \\
 7&0.016591966498537 \\
 8&0.013132015183494 \\
 9&0.010684547791392 \\
10&0.008884576737074 \\
11&0.007519096948920 \\
12&0.006456674647995 \\
\hline
\end{tabular}
\end{center}
\end{table}

\end{itemize}

\subsubsection{Analysis of the data.}
\label{sec7.2.2}
We will now use the TM data to check the theoretical predictions.
We expect that the error induced by the error on $\chi$ given in 
Tables \ref{latA} and \ref{latB} is small compared to that due to the 
neglected higher-order corrections in (\ref{eq5.2}). Therefore, 
instead of performing a fit, we considered as 
many data points as the number of free parameters 
of the Ansatz, and then required  
the Ansatz to be exact for them.  
This gives a system of equations that is then solved for the
free parameters. 
We always used consecutive values of $L$, i.e. $L_1=L$, 
$L_2=L-1$,...,$L_n=L-n+1$, where $n$ is the number of free parameters.
Errors were estimated from the variation of the results with the lattice
size and by comparison of different Ans\"atze.

As a preliminary test we checked that $y=-2$ for 
the leading correction to scaling.
For this purpose we studied the Ansatz
\be
\label{checky}
\chi_L(0,0) = L^{7/4} \left(c_0 + c_1 L^{y} \right) + D_0,
\ee
with $c_0$, $c_1$, and $y$ as free parameters. 
The results are summarized in Table
\ref{yanalysis}.
\begin{table}
\caption{\sl
Numerical results from the Ansatz (\ref{checky}) in 
geometries (A) and  (B).}
\vskip0.5cm
\label{yanalysis}
\begin{center}
\begin{tabular}{|r|r|r|r|}
\hline
\multicolumn{4}{|c|}{Geometry (A)}\\
\hline
\multicolumn{1}{|c|}{$L$}&
\multicolumn{1}{|c|}{$c_0$}&
\multicolumn{1}{|c|}{$c_1$}&
\multicolumn{1}{|c|}{$y$} \\
\hline
12 & 1.0919299 & --0.0964 & --2.102 \\
13 & 1.0919370 & --0.0915 & --2.076 \\
14 & 1.0919414 & --0.0881 & --2.057 \\
15 & 1.0919441 & --0.0857 & --2.044 \\
16 & 1.0919460 & --0.0838 & --2.034 \\
17 & 1.0919472 & --0.0823 & --2.026 \\
\hline
\hline
\multicolumn{4}{|c|}{Geometry (B)}\\
\hline
\multicolumn{1}{|c|}{$M$}&
\multicolumn{1}{|c|}{$c_0$}&
\multicolumn{1}{|c|}{$c_1$}&
\multicolumn{1}{|c|}{$y$} \\
\hline
  8 & 1.0919297 &  0.0689  &--1.922 \\
  9 & 1.0919388 &  0.0720  &--1.946 \\
 10 & 1.0919435 &  0.0743  &--1.962 \\
 11 & 1.0919461 &  0.0761  &--1.973 \\
 12 & 1.0919477 &  0.0775  &--1.982 \\
\hline
\end{tabular}
\end{center}
\end{table}
For both geometries the numerical result for $y$ 
approaches $-2$ as $L$ increases.
For our largest lattice sizes, the deviation from $-2$ is about $1\%$.
In the following analysis we shall assume $y=-2$. 

Next we analyzed the data with \reff{eq5.2}.
For geometry (A), by using the known values of $D_0$, $d_0$, and $d_2$, 
we found 
\begin{eqnarray}
c_0^A &=& 1.09195056(4) \\
c_1^A &=&-0.07914(5), 
\end{eqnarray}  
where the quoted uncertainties were obtained by comparing the 
results of the Ansatz \reff{eq5.2} with those obtained by adding 
$c_3$ as a free parameter.

For geometry (B), by using the known value of $D_0$, we obtain 
\begin{eqnarray}
c_0^B &=& 1.0919504(2) \\
c_1^B &=& 0.0794(4), \\
d_0^B &=& 0.019(5).
\end{eqnarray}
Our predictions \reff{prediction-chiL} appear to be very well satisfied. 
Moreover, our result for $c_0$ is in good agreement with, although much more 
precise than, the estimate\footnote{We
report here the result of their fit with $\Delta=7/4$, since this is the 
correct theoretical behavior.}
of \cite{ssv1}, $c_0 = 1.09210(11)$.

If we assume $d_0^B = d_0^A$ and use \reff{C1A-theory}, we obtain the 
more precise estimate 
\begin{eqnarray}
c_0^B &=& 1.0919506(2)  \\
c_1^B &=& 0.0790(2),
\end{eqnarray}
where the error was obtained by comparing the results with and without
the parameter $d_2$.

{} From the above analysis we see that, within the errors, the
coefficients of the  $1/L^2$ correction are equal in magnitude and opposite in
sign for the two geometries. Since the two lattices are rotated by $\pi/4$
this implies that this correction {\sl is completely due
to the spin-four operator $T^2 +\bar T^2$} and 
that the scalar operator $T\bar T$
is absent, in agreement with the conjecture (d0).

\section{Concluding remarks and open issues.}
\label{sec8}

In this paper we have discussed the presently available results
for the corrections to scaling in the two-dimensional Ising model.
We have shown that all results are in agreement with the 
RG and CFT predictions. The only missing point here is a complete 
analysis of the RG resonances and consequently an 
extension of the scaling forms \reff{F-RG} and \reff{F-RG-finiteL}
to take into account the logarithmic structure found in \cite{g2000}.
We have also shown that the existence of an exact symmetry in the lattice 
models that maps the high-temperature phase onto the low-temperature 
one plays a very important role and explains the symmetry 
properties of the results.

However, the lattice Ising model shows also features that are {\em not} 
predicted by CFT and RG and that can be explained if some 
additional conjectures are made. A list of them is reported in 
Sec. \ref{sec4.1}. Let us summarize the evidence we have: 
\begin{itemize}
\item Conjectures (a) and (b). They allow to explain the 
symmetry properties under $\tau\to-\tau$ of the free energy and of 
its derivatives for $h=0$. Further evidence may be obtained by 
analyzing $\chi$ on the triangular lattice and checking 
whether the functions $\widehat{F}_\pm(\tau)$ are even in $\tau$.

\item Conjecture (c1): The functions $\tilde{f}_\pm$ do not depend on the 
$[\sigma]$-family fields. This is supported by the exact known 
results for $F(\tau,0)$ and $M(\tau,0)$ and by the results of 
\cite{g2000}. Further evidence is obtained from the absence
of a  leading logarithmic correction in 
higher-point correlation functions \cite{CHPV-gstar,CHPV-eqst}.

\item Conjecture (c2): The functions $\tilde{f}_\pm$ are constants 
(this is the original conjecture of \cite{af}). The 
independence of $\tilde{f}_\pm$ from the $\sZ_2$-even
scaling fields is supported by the  finite-size results 
of \cite{Salas-01,ih-00-square} discussed in Sec. \ref{sec6}. 
The conjecture follows from this observation and the conjecture (c1).
Conjecture (c2), together with the conjectured formula \reff{eq2:What} 
can be further checked by computing the
logarithmic term(s) in $\partial^n F/\partial \tau^n$ at the critical
point for $n>2$.

\item Conjecture (d0): The nonlinear scaling field of $T\bar{T}$ 
vanishes at the critical point. 
On the square lattice we have ample evidence 
in favor of (d0), which is the only conjecture needed to explain 
the existing results. Indeed, it is supported by: 
\begin{itemize}
\item[(1)] The infinite-volume results of \cite{g2000}.
\item[(2)] The behavior of $\xi(\theta)$ discussed in Sec. \ref{sec5.1}.
\item[(3)] The dependence of $\chi$ at the critical point from the 
boundary conditions, see Sec. \ref{sec7}. 
\item[(4)] The behavior of the two-point function at the critical point,
see \cite{CGM-01}.
\item[(5)] The behavior of the free energy on the critical isotherm,
see \cite{CaHa99}.
\end{itemize}
Moreover, all triangular-lattice results are compatible with it. For these 
reasons, we believe it is more than a conjecture and it is essentially
proved. It is interesting to notice that a similar cancellation 
is observed in the finite-size scaling behavior of the free energy and 
of the mass spectrum in the one-dimensional Ising quantum chain,
see \cite{r87a}. 

\item Conjecture (d1): The operator $T\bar{T}$ is decoupled. 
We have evidence for the validity of this conjecture in the triangular-lattice
Ising model.  The analysis of the 
correlation length $\xi(\theta)$ on the triangular lattice
shows that $\mu_{T\bar{T}}(0)$ vanishes at least up
to terms of order $O(\tau^6)$. There are several calculations that should 
be feasible and would add further support to the validity of (d1) on the 
triangular lattice:
\begin{itemize}
\item[(1)] The extension of the results of Ref. \cite{g2000} to the 
triangular lattice.
\item[(2)] The study of the dependence on the boundary conditions
of the observables studied in Sec. \ref{sec6} at the critical point on
the triangular lattice. This would unambiguously identify the 
spin of the leading irrelevant operator.
\item[(3)] The study of $\chi$ at the critical point on a triangular 
lattice. It is particularly important to verify whether $d_3$,
cf. \reff{sviluppochiL-secondo},  vanishes or not.  
If it does, it provides the only available evidence 
for $\lambda_{T\bar{T}}(0) = 0$, and thus it 
would strengthen the conjecture.
\end{itemize}

\item Conjecture (d2): Only nonzero-spin operators are present. 
We have evidence for this conjecture on the triangular lattice.
The absence of spin-zero operators beside $T\bar{T}$ is based on the 
results of Sec. \ref{sec5.2} and \ref{sec6} where we showed that the 
existing exact results require $\mu(0)=0$ for the spin-zero 
identity-family operator $Q^I_4 \bar{Q}^I_4$ with $y=-6$.
The studies (1) and (2) 
mentioned at the previous point would further check the conjecture.
In particular, they can verify whether 
$v(0)=0$ for the spin-zero 
$[\sigma]$-family operator $Q_3^\sigma \bar{Q}_3^\sigma$ 
with $y=-4-\frac18$. 
\end{itemize}

Of course, as they stand, these conjectures are just ``ad hoc'' prescriptions, 
whose only merit is that of providing an economical way to explain 
all existing results.
It would be very important to understand if there is some symmetry argument 
that could explain them. 

There remain several open questions. First of all, one may ask whether
these conjectures apply to the 
nearest-neighbor Ising model on any regular lattice or whether 
some of them depend on the lattice structure.
Another important question is how important the nearest-neighbor
condition is: Do some of these conjectures apply also to the 
Ising model with extended interactions? Finally, one may ask whether 
these cancellations are also observed in other models.
Concerning this last question, we should mention the results of~\cite{grv87}
for the three-state Potts quantum chain, which were 
compared with the CFT predictions in~\cite{r87a}. Again, the
energy-momentum tensor contribution turns out to be compatible with zero.
However, at variance with the Ising case, there is here, 
at next-to-leading order,
a clear signature of a finite-size correction due to a {\sl
scalar} irrelevant operator.  Even if the Potts case
is slightly different from the Ising one,
since this irrelevant operator is
actually a {\sl primary} operator 
(more precisely is the one with conformal weight
$h=\frac{7}{5}$), this result indicates that our conjecture (d2), if true,
is specific of the Ising model and could be somehow related to the fact that 
the model is soluble on the lattice. On the other hand, the vanishing of the
correction due to the energy-momentum tensor seems to be a more 
general phenomenon. In order to understand the validity of (d0), it would 
be interesting to extend these analyses to the generic $q$-state Potts model
or to other specific values of $q$ (for instance, to percolation).

\vskip 1cm
{\bf  Acknowledgements.}
We thank Malte Henkel for several useful suggestions.

\end{document}